\documentclass[12pt,letterpaper]{article}
\usepackage[dvips]{graphicx,color}
\usepackage{amsmath,amssymb}
\usepackage[dvips,letterpaper,text={6.5in,9in}]{geometry}
\usepackage{fancyhdr}
\usepackage{verbatim}

\newcommand\ltap{\
  \raise.3ex\hbox{$<$\kern-.75em\lower1ex\hbox{$\sim$}}\ } 
\newcommand\gtap{\
  \raise.3ex\hbox{$>$\kern-.75em\lower1ex\hbox{$\sim$}}\ } 
\newcommand\simge{\mathrel{%
   \rlap{\raise 0.511ex \hbox{$>$}}{\lower 0.511ex \hbox{$\sim$}}}}
\newcommand\simle{\mathrel{
   \rlap{\raise 0.511ex \hbox{$<$}}{\lower 0.511ex \hbox{$\sim$}}}}

\newcommand{\slashchar}[1]%
        {\kern .25em\raise.18ex\hbox{$/$}\kern-.75em #1}
\def\lsim{\mathrel{\raise.3ex\hbox{$<$\kern-.75em\lower1ex\hbox{$\sim$}}}}
\def\gsim{\mathrel{\raise.3ex\hbox{$>$\kern-.75em\lower1ex\hbox{$\sim$}}}}
\newcommand{\bs}{\boldsymbol}

\newcommand\CG{{\cal G}}
\newcommand\CH{{\cal H}}

\newcommand\CO{{\cal O}}

\newcommand\CW{{\cal W}}

\newcommand\be{\begin{equation}} 
\newcommand\ee{\end{equation}} 
\newcommand\bea{\begin{eqnarray}}
\newcommand\eea{\end{eqnarray}}
\newcommand\ba{\begin{array}}
\newcommand\ea{\end{array}}
\newcommand\nn{\nonumber}
\newcommand\whW{\widehat W}
\newcommand\wtW{\widetilde W}

\newcommand\homega{\widehat\omega_{Da}}
\newcommand\whomega{\widehat\omega_8}
\newcommand\whomegaz{\widehat\omega_{8,0}}
\newcommand\dtwoE{d^2E(W)/d\Lambda^2}

\newcommand{\half}{\ensuremath{\frac{1}{2}}}

\newcommand\ts{\thinspace}
\newcommand\ra{\rightarrow}

\newcommand\Lra{\Longrightarrow}

\newcommand\ol{\bar}

\newcommand\gev{{\rm GeV}}
\newcommand\tev{{\rm TeV}}

\newcommand\lvac{\langle \Omega \vert}
\newcommand\rvac{\vert \Omega \rangle}

\begin{document}
\title{
\vskip -15mm
\begin{flushright}
\vskip -15mm
{\small BUHEP-05-1\\
hep-ph/0501204\\}
\vskip 5mm
\end{flushright}
{\Large{\bf Accidental Goldstone Bosons}}\\
}
\author{
{\large Kenneth Lane\thanks{lane@bu.edu}\,\, and Adam
  Martin\thanks{aomartin@bu.edu}}\\
{\large {$$}Department of Physics, Boston University}\\
{\large 590 Commonwealth Avenue, Boston, Massachusetts 02215}\\
}
\maketitle

\begin{abstract}
  We study vacuum alignment in theories in which the chiral symmetry of a set
  of massless fermions is both spontaneously and explicitly broken. We find
  that transitions occur between different phases of the fermions' CP
  symmetry as parameters in their symmetry breaking Hamiltonian are varied.
  We identify a new phase that we call pseudoCP-conserving. We observe first
  and second-order transitions between the various phases. At a second-order
  (and possibly first-order) transition a pseudoGoldstone boson becomes
  massless as a consequence of a spontaneous change in the discrete, but not
  the continuous, symmetry of the ground state. We relate the masslessness of
  these ``accidental Goldstone bosons'' (AGBs) bosons to singularities of the
  order parameter for the phase transition. The relative frequency of
  CP-phase transitions makes it commonplace for the AGBs to be light, much
  lighter than their underlying strong interaction scale. We investigate the
  AGBs' potential for serving as light composite Higgs bosons by studying
  their vacuum expectation values, finding promising results: AGB vevs are
  also often much less than their strong scale.

\end{abstract}

%%%%%%%%%%%%%%%%%%%%%%%%%%%%%%%%%
%%%%%%%%%%%%%%%%%%%%%%%%%%%%%%%%%

\newpage

\section*{I. Introduction}

In this paper we describe phenomena we believe to be very general but which
appear to have received almost no attention in particle physics.\footnote{The
  principal exception is a brief passage in Dashen's classic paper on vacuum
  alignment~\cite{Dashen:1971et}; see the $m_\eta^2$ discussion at the end of
  his section III.} These are the presence of
various phases of CP symmetry, of transitions among these phases, and of
anomalously light bosons which become massless at these phase transitions.
While, in the model calculations we present, the massless state is a
pseudoGoldstone boson (PGB) of an approximate chiral symmetry, the boson's
masslessness is {\em not} due to the restoration of its associated continuous
chiral symmetry.  Rather, it is due to a change in the phase of the discrete
CP symmetry.  Following Dashen, who first observed this phenomenon in the
context of QCD, we call these ``accidental Goldstone bosons'' (AGBs). Unlike
Dashen, however, we do not believe the AGB's mass necessarily is restored by
higher-order corrections.  Rather, we suspect that corrections only shift the
values of parameters at which phase transitions occur.

This study grew out of earlier work on vacuum alignment in technicolor
theories of dynamical electroweak symmetry
breaking~\cite{Lane:2000es,Lane:2002wv}; also see Ref.~\cite{Martin:2004ec}
for a recent summary. However, although our calculations are similar to those
used in technicolor, we are confident our conclusions extend beyond that
setting. Some of the phenomena we observe also occur in QCD when one allows
an odd number of real quark masses to become negative (so that $\bar \theta =
\pi$)~\cite{Dashen:1971et,Creutz:2003xu}). We see no reason they would not
also occur in models different from the type we investigate here. They may
even have relevance to condensed matter systems.

The model we use assumes $N$ massless Dirac fermions $T_i = (T_{Li},T_{Ri})$,
$i=1,\dots, N$, transforming according to a complex representation of a
strongly-coupled $SU(N_c)$ gauge group. The fermions' chiral flavor symmetry,
$G_f = SU(N)_L \otimes SU(N)_R$, is spontaneously broken to an $SU(N)$
subgroup. It is convenient to work in a ``standard vacuum'' $\rvac$ whose
symmetry is the vectorial $S_f = SU(N)_V$ defined by the $SU(N_c)$-invariant
condensates\footnote{We work in vacua with the instanton angle $\theta_c$
  rotated to zero. Condensates are assumed to be CP-conserving.}
\be\label{eq:standard}
\lvac \ol T_{Li} T_{Rj} \rvac = \lvac \ol T_{Ri} T_{Lj} \rvac =
-\delta_{ij} \Delta_T \,.
\ee
The condensate $\Delta_T$ is renormalized at the $SU(N_c)$ scale $\Lambda_T$,
which (for $T_i \in {\bs N_c}$ of $SU(N_c)$) is commonly assumed to be
$\Lambda_T \simeq 4\pi F_\pi$ and, then, $\Delta_T \simeq 2\pi F^3_\pi$.
Here, $F_\pi$ is the decay constant of the massless Goldstone bosons,
$\pi_a$, $a=1,\dots,N^2-1$, resulting from the spontaneous chiral symmetry
breaking. It is normalized by the relation $\lvac \bar T \gamma_\mu \gamma_5
t_a T |\pi_b(p)\rangle = i F_\pi p_\mu \delta_{ab}$, with $t_a = \lambda_a/2$
so that ${\rm Tr}(t_a t_b) = \half \delta_{ab}$.

The chiral $SU(N)_L \otimes SU(N)_R$ symmetry is also broken explicitly by
the $SU(N_c)$-invariant four-fermion interactions
\be\label{eq:Hprime0}
\CH' = \sum_{ijkl} \Lambda_{ijkl} \ol T_{Li} \gamma^\mu T_{Lj}
 \ol T_{Rk} \gamma_\mu T_{Rl} + {\rm LL\,,\,RR\,\,\,terms} + {\rm h.c.}\,,
\ee
where the unexhibited LL and RR terms are irrelevant for our further
discussion. The $\Lambda_{ijkl} = \Lambda^*_{jilk}$ are inverse squared
masses of gauge bosons (or scalars) exchanged between the ``currents'' (or
their Fierz transforms) in $\CH'$. They are chosen in numerical calculations
so that {\em all} $\pi_a$-symmetries are explicitly broken.  Ordinarily,
then, we would expect the PGBs to acquire positive mass-squared.  Finally, we
assume that $\CH'$ is T-invariant, i.e.,
\be\label{eq:Tinv}
\Lambda_{ijkl} = \Lambda^*_{ijkl}\,.
\ee

Now, there may be a mismatch between the standard vacuum $\rvac$ and the one
in which $\CH'$ of Eq.~(\ref{eq:Hprime0}) gives positive mass-squared to all
$\pi_a$. It is therefore necessary to ``align the vacuum'', more precisely,
to determine the correct ground state $|{\rm vac}\rangle$ of the
theory~\cite{Dashen:1971et}. In this state, $\langle{\rm vac}|\CW^{-1} \CH'
\CW|{\rm vac}\rangle$, varied over $\CW \in SU(N)_L \otimes SU(N)_R)$, is a
minimum at $\CW = 1$. We follow Dashen's procedure, which is based on
lowest-order chiral perturbation theory and minimize the vacuum energy
defined over the infinity of perturbative ground states
$|\Omega(\CW)\rangle$:
\bea\label{eq:vacE}
E(W) &=&  \langle\Omega(\CW)| \CH' |\Omega(\CW)\rangle \equiv
\lvac \CW^{-1} \CH' \CW\rvac \nn\\
&=& - \sum_{ijkl}\Lambda_{ijkl} W_{jk} W^\dag_{li} \Delta_{TT} + {\rm constant}\,.
\eea
Here, we used (since $T_{L,R\, i}$ transform as a complex representation of
$SU(N_c)$)
\be\label{eq:deltaTT}
\lvac \ol T_{Li} \gamma^\mu T_{Lj}
 \ol T_{Rk} \gamma_\mu T_{Rl} \rvac = - \delta_{il}\delta_{jk} \Delta_{TT}\,.
\ee
Since $\rvac$ is invariant under $SU(N)_V$ transformations, the $SU(N)$
matrix $W = W_L W_R^\dag \in G_f/S_f$ is the only physically meaningful
combination of $W_L$ and $W_R$. The four-fermion condensate is renormalized
at the scale of the exchanged-boson masses making up the $\Lambda_{ijkl}$.
This scale is likely to be much greater than $\Lambda_T$.  In a QCD-like
$SU(N_c)$ theory, $\Delta_{TT} \sim \Delta_T^2$. If $SU(N_c)$ is a walking
gauge
theory~\cite{Holdom:1981rm,Appelquist:1986an,Akiba:1986rr,Yamawaki:1986zg},
$\Delta_{TT}$ could be much larger than $\Delta_T^2$, partially overcoming
the suppression by the $\Lambda_{ijkl}$.

Note that CP-invariance of $\CH'$ implies that $E(W) = E(W^*)$. The $W_0$
which minimizes $E$ is defined up to a factor $Z_N^m = \exp{(2im\pi/N)}$,
$m=0,\dots,N-1$. If, apart from this trivial ambiguity, $W_0 \neq W_0^*$, then
the correct chiral-perturbative ground state is discretely degenerate and CP
symmetry is spontaneously broken. Equivalently, and more conveniently, the
Hamiltonian $\CH'(\CW_0) = \CW_0^{-1} \CH'\, \CW_0$ correctly aligned with
the standard $\rvac$ violates CP.

It is convenient to make the $SU(N)_V$ transformation $W_{L,R} \ra W_{L,R}
W^\dag_R$. This amounts to computing $\CH'(\CW_0)$ with $W_L = W$, $W_R = 1$.
Then, dropping the subscript ``0'' from now on,
\bea\label{eq:HprimeW}
\CH'(W) &=& \sum_{ijkl} \Lambda^W_{ijkl} \ol T_{Li} \gamma^\mu T_{Lj}
 \ol T_{Rk} \gamma_\mu T_{Rl} + {\rm LL\,,\,RR\,\,\, terms}+ {\rm h.c.}\,,\nn\\
\Lambda^W_{ijkl} &=& \sum_{i',j'} \Lambda_{i'j'kl} W^\dag_{ii'} W_{j'j}\,.
\eea
After this vectorial transformation, the PGB mass-squared matrix is still
calculated using the axial charges, formally defined by $Q_{5a} = \int d^3x\,
T^\dag \gamma_5 t_a T$. To lowest order in chiral perturbation
theory~\cite{Dashen:1969eg},
\be\label{eq:Msq}
F_\pi^2 M^2_{ab} = i^2 \left\langle\Omega\left| \left[Q_{5a}, 
\left[Q_{5b},\CH'(W)\right]\right]\right|\Omega\right\rangle\,.
\ee
For the Hamiltonian in Eq.~(\ref{eq:HprimeW}).
\bea\label{eq:Msqtwo}
F_\pi^2 M^2_{ab} &=&
2 \sum_{ijkl} \Lambda_{ijkl} \Bigl[
\left(\left\{t_a, t_b\right\}W^\dag\right)_{li} \, W_{jk} +
W_{li}^\dag\, \left(W\left\{t_a, t_b\right\}\right)_{jk} 
\nn\\
& &\qquad -2 \left(t_a W^\dag\right)_{li} \,
\left(W t_b \right)_{jk} 
-2 \left(t_b W^\dag\right)_{li} \,
\left(W t_a\right)_{jk}\ts \Bigr]\Delta_{TT} \,.
\eea
Finally, it is very useful to parameterize $W$ in the form
\be\label{eq:Wform}
W = D_L K D_R \ts.
\ee
Here, $D_{L,R}$ are diagonal $SU(N)$ matrices, each involving $N-1$
independent phases $\chi_{L,R\, i}$, and $K$ is an $(N-1)^2$--parameter CKM
matrix which may be written in the standard Harari-Leurer
form~\cite{Harari:1986xf}.

The remainder of this paper is organized as follows: In Sec.~II we review
vacuum alignment for the model we've described. The four-fermion form of
$\CH'$ implies a linking of the phases in $W$ which allows the possibility of
three ``phase phases'' with different CP properties. We call these three
phases CP-conserving (CPC), pseudoCP-conserving (PCP), and CP-violating
(CPV). In the CPC phase, $W$ is $Z_N^m$ times a real matrix and $\CH'(W)$ is
real. In the PCP phase, $W$ is not simply $Z_N^m$ times a real matrix, but
the phases in $W$ are {\em rational} multiples of $\pi$ and the CKM matrix
$K$ is real. The phases in $\CH'(W)$ are also rational, but the Hamiltonian
is not merely real up to an overall phase. However, introducing the aligning
matrix $\whW = D_R W D_R^\dag = D_R D_L K$, we show that the LR terms in
$\CH'(\whW)$ are real, i.e., CP-conserving, in {\em both} the CPC and PCP
phases. In the CPV phase, the phases of $W$ are not rational multiples of
$\pi$, $K$ is not real, and $\CH'(W)$ is definitely CP-violating.

We carry out vacuum alignment numerically in a three-flavor ($SU(3)$) model,
varying one $\Lambda_{ijkl} \equiv \Lambda$ in $\CH'$. We observe each of
these CP phases and note that the transitions between them are either first
or second order --- defined here as whether the first or second derivative of
$E(W)$ with $\Lambda$ is discontinuous. Although we are varying just one of
the parameters in $\CH'$, it is obvious that the phase transitions occur on
surfaces in the space of $\Lambda_{ijkl}$'s. Our calculations are merely
along a single trajectory in this $\Lambda$-space. There are two PGBs whose
$M^2$ is much less than those of the other six. These are the accidental
Goldstone bosons of this model. At all second-order (and, apparently,
first-order) transitions one of these light PGBs becomes massless.  We
explain why this happens.\footnote{Most of the features of vacuum alignment
  described in Sec.~II and this $SU(3)$ model were discussed in
  Ref.~\cite{Lane:2000es}.  The treatment in the present paper is much more
  incisive.} Our calculations indicate that light AGBs are commonplace, at
least as long as the $\Lambda_{ijkl}$ are the same order of magnitude. Then,
competition among the $\Lambda$'s means that one is never very far from a CP
phase transition and a surface in $\Lambda$-space on which an AGB mass
vanishes.

Sections~III and~IV are devoted to understanding the phase transitions in
more depth. In Sec.~III we present a remarkable formula for $d^2
E(W)/d\Lambda^2$ which connects the vanishing of $M^2$ to singular behavior
of the ``diagonal phases'' of $W$, the $N-1$ phases $\omega_{Da}$, $a=n^2-1 =
3,8,\dots,N^2-1$, of $W$ in its diagonal form. This formula also directly
relates the AGBs to the diagonal phases. The formula is derived in
Appendix~A. An analytic example of how it works is given in Appendix~B using
Dashen's model --- three quarks with negative masses. We also illustrate it
numerically for the $SU(3)$ model.  In Sec.~IV we focus on the aligning
matrix $\whW$. In the CPC and in what we call PCP-1 phases, $\whW =
e^{im\pi/N} \wtW$ where $\wtW$ is real. In PCP-2 phases, $\whW$ cannot be
written this way. In the CPC phase of the $SU(3)$ model we study, $\wtW$ {\em
  appears} to be symmetric.\footnote{As we discuss in Sec.~IV, this is very
  nearly true numerically, but it appears to be an artifact of how we chose
  the model's $\Lambda_{ijkl}$.} It is shown that this implies the normalized
diagonal phases $\homega = \omega_{Da}/\sqrt{n(n-1)/2}$ are rational
multiples of $\pi$. In PCP-1 phases, $\wtW$ is not symmetric. In this case,
some but not all the $\homega$ are rational. We spell out the conditions for
determining how many $\homega$ are rational. In the PCP-2 phase, none of the
$\homega$ are rational. This is startling since all the phases in $W$ are.

Finally, in Sec.~V we discuss one potential application of AGBs: light
composite Higgs bosons for electroweak symmetry
breaking~\cite{Kaplan:1983fs,Kaplan:1983sm}. A light composite Higgs boson is
a bound state whose mass {\em and} vacuum expectation value (vev) are {\em
  naturally} much less than the energy scale at which its binding occurs. The
effort to construct realistic models of light composite Higgses has been
driven by the strong experimental evidence in favor of the standard model
with a light Higgs boson. Recently, much of this effort has focused on the
little Higgs scenario~\cite{Arkani-Hamed:2001nc,Arkani-Hamed:2002qy,
  Arkani-Hamed:2002qx,Schmaltz:2002wx}. Little Higgs bosons are PGBs that are
anomalously light because interlocking {\em continuous} symmetries need to be
broken by several weakly-coupled interactions, making their nonzero mass a
multiloop effect. In most models so far, little Higgses acquire masses in two
loops so that a compositeness scale of $\Lambda_{lH} \simeq 4\pi F_{lH}
\simeq 10\,\tev$ yields a mass and vev of $M_{lH} \simeq 100\,\gev$ and
$v_{lH} = 100$--$200\,\gev$.

Accidental Goldstone bosons can easily have $M \ll \Lambda_T \simeq
4\pi F_\pi$, the $T$-fermion scale. The challenges are (1) a vev $v \simeq M
\ll \Lambda_T$, (2) embedding the AGB structure into electroweak
$SU(2)\otimes U(1)$ symmetry, and (3) coupling the AGBs to quarks and leptons
to account for their masses and mixings (without running afoul of
flavor-changing neutral current and precision electroweak constraints). In
Sec.~V, we study the first of these, the magnitudes of the AGB vevs, and
find that they too are often much smaller than $\Lambda_T$.

\section*{II. Vacuum Alignment and the Phase Phases}

There are several useful forms of the alignment matrix $W$:
\be\label{eq:Wforms}
W_{ij} \equiv (D_L K D_R)_{ij} = e^{i(\chi_{Li} + \chi_{Rj})} K_{ij} =
 |W_{ij}| e^{i\phi_{ij}} \,.
\ee
There are $N^2$ $\phi_{ij}$. The CKM matrix $K$ has $\half N(N-1)$ angles
$\theta_{ij}$ ($1\le i < j \le N$) and $\half (N-1)(N-2)$ phases $\chi_{ij}$
($1\le i < j-1 \le N-1$).

It was shown in Ref.~\cite{Lane:2000es} that there are three possibilities
for the phases $\phi_{ij}$. Consider an individual term, $-\Lambda_{ijkl} \,
W_{jk} \, W^\dag_{li} \, \Delta_{TT}$, in $E(W)$. If $\Lambda_{ijkl} > 0$,
this term is least if $\phi_{il} = \phi_{jk}$; if $\Lambda_{ijkl} < 0$, it is
least if $\phi_{il} = \phi_{jk} \pm \pi$. Thus, $\Lambda_{ijkl} \neq 0$ links
$\phi_{il}$ and $\phi_{jk}$, and tends to align (or antialign) them. However,
the constraints of unitarity may partially or wholly frustrate this
alignment. This then gives the three phase phases:

\begin{itemize}
  
\item[1.] All $\phi_{ij}$ are linked to one another and unitarity allows them
  to be equal.  Unimodularity of $W$ implies all $\phi_{ij} = 2m\pi/N$ (mod
  $\pi$) for fixed $m=0,\dots,N-1$. Then $W = Z_N^m$ times a real orthogonal
  matrix, and all the terms in $\CH'(W)$ are real. This is the CPC phase.
  
\item[2.] Not all $\phi_{ij}$ are linked to one another. Still, if unitarity
  allows it, the $\phi_{ij}$ are again rational multiples of $\pi$, but
  generally not equal to one another (mod $\pi$). Rather, their values are
  various multiples of $\pi/N'$ for one or more integers $N'$. As explained
  in Ref.~\cite{Lane:2000es}, $K$ is real and this is a necessary condition
  for rational phases. We also showed there that, while $\CH'(W)$ is not
  real, the phases in the $\Lambda^W_{ijkl} = \Lambda_{i'j'kl} W^\dag_{ii'}
  W_{j'j}$ are rational. Thus, we call this the PCP phase. We repeat the
  proof: If $K$ is real and $\Lambda_{i'j'kl} \neq 0$ then $\phi_{j'k}$ and
  $\phi_{i'l}$ are linked and, in this phase, $\phi_{j'k} - \phi_{i'l} =
  \chi_{Lj'} + \chi_{Rk} - \chi_{Li'} - \chi_{Rl} = 0$ (all phase equalities
  are mod~$\pi$). The phase of an individual term in the sum for
  $\Lambda^W_{ijkl}$ is then $\phi_{j'j} - \phi_{i'i} = \chi_{Lj'} +
  \chi_{Rj} - \chi_{Li'} - \chi_{Ri} = \chi_{Rj} - \chi_{Ri} + \chi_{Rl} -
  \chi_{Rk}$. This is a rational phase which is the same for all terms in the
  sum over $i', j'$. Indeed,
\be\label{eq:LamW}
\Lambda^W_{ijkl} = e^{i(\chi_{Rj} - \chi_{Ri} + \chi_{Rl} - \chi_{Rk})}
\sum_{i',j'}|\Lambda_{i'j'kl} K_{i'i} K_{j'j}|\,\, {\rm sgn}(K_{i'l} K_{i'i}
  K_{j'k}K_{j'j})\,.
\ee
We see from Eqs.~(\ref{eq:HprimeW},\ref{eq:LamW}) that the vectorial change
of variable $T_{L,R} = D_R^\dag T'_{L,R}$ makes all the LR terms real in
Eq.~(\ref{eq:LamW}). Under this transformation, the aligning matrix $W$
becomes $\whW = D_R W D_R^\dag = D_R D_L K$ and, of course, $E(\whW) = E(W)$.
Although the LR terms are made real by this transformation, the LL and RR
terms generally are not because there is no phase-linking argument for the
$\Lambda^{LL,RR}_{ijkl}$. Whether they have rational phases or not is a
model-dependent (and $W$-convention-dependent) question.

\item[3.] Whether or not the $\phi_{ij}$ are linked, unitarity frustrates
  their alignment so that they are all unequal, irrational multiples of
  $\pi$, random except for the constraints of unitarity and unimodularity.
  This is the CPV phase in which the phases in $\CH'(W)$ are irrational hash.

\end{itemize}

A demonstration of these three phases is provided by a model with three
flavors.\footnote{This model was studied in Ref.~\cite{Lane:2000es}, but only
  over the range $\Lambda = 0.5$--1.1. The phase transitions near $\Lambda
  =1.9$ and~2.8 were missed in that discussion.} The chiral symmetry
$SU(3)_L\otimes SU(3)_R$ is broken in the vacuum to $SU(3)$. The model's
eight Goldstone bosons get mass from a Hamiltonian $\CH'$ with nonzero
couplings
\bea\label{eq:SU3model}
&&\Lambda_{1111} = \Lambda_{1221} = \Lambda_{2112} =
  \Lambda_{1212} = \Lambda_{2121} = 1.0 \nn\\
&&\Lambda_{1122} = 1.5 \ts, \quad \Lambda_{1133} = 1.4 \nn\\
&&  \Lambda_{1331} = \Lambda_{3113} = 1.6, \quad
\Lambda_{1313} = \Lambda_{3131} = 1.8 \nn\\
&&\Lambda \equiv \Lambda_{1222} = \Lambda_{2122} = \Lambda_{2212} =
\Lambda_{2221} = 0.0-3.0
\eea
These tend to align $\phi_{11} = \phi_{22} = \phi_{33} = \phi_{12} =
\phi_{21}$ and $\phi_{13} = \phi_{31}$. The phases $\phi_{23}$ and
$\phi_{32}$ are not linked by these $\Lambda$'s. 

\begin{figure}[t]
   \begin{center}
     \includegraphics[width=3.75in, height = 5.5in, angle=270]{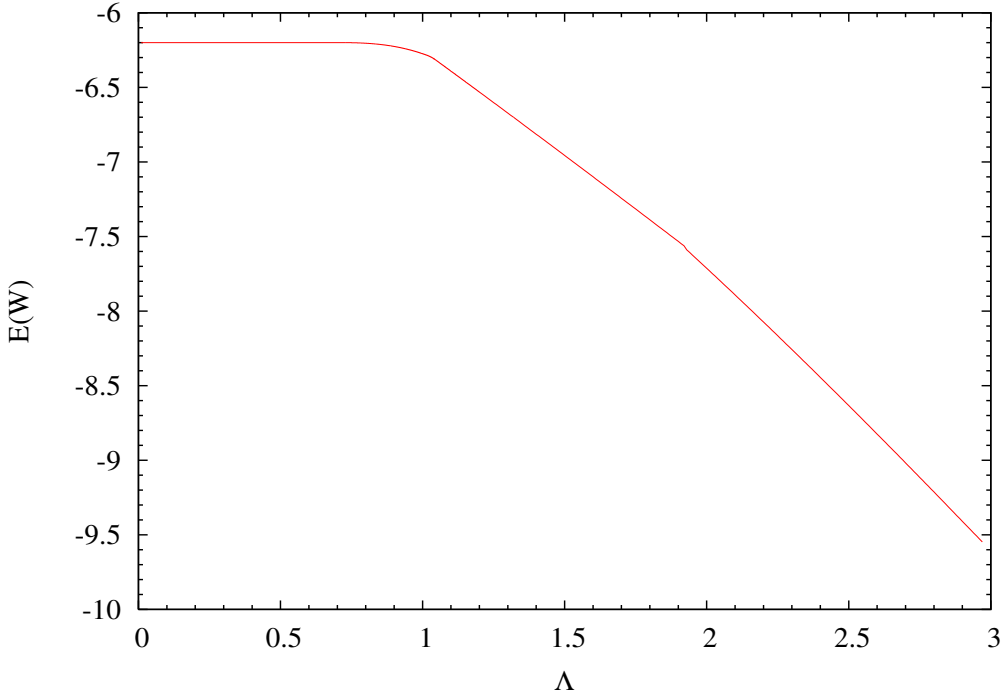}
     \caption{The vacuum energy $E(W)$ in the $SU(3)$ model as a function of
       $\Lambda = \Lambda_{1222} = \Lambda_{2212}$. Note the discontinuous
       slope near $\Lambda = 1.9$. }
     \label{fig:AGB_fig_1}
   \end{center}
 \end{figure}

\begin{figure}[t]
   \begin{center}
     \includegraphics[width=3.65in, height = 5.5in, angle=270]{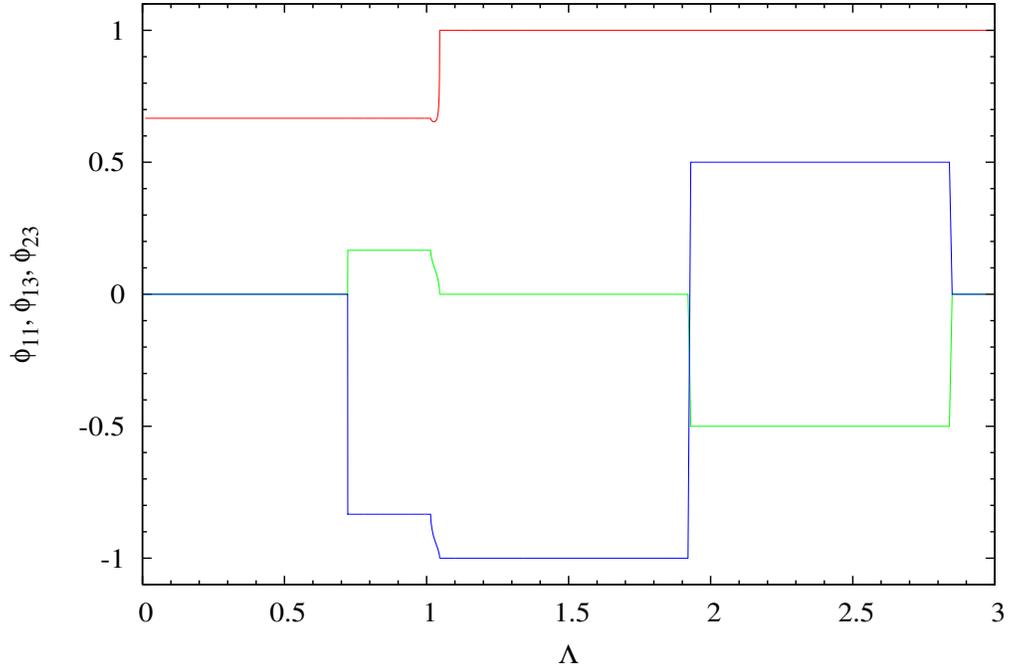}
     \caption{The $W$-phases $\phi_{11}/\pi$ (red), $\phi_{13}/\pi$ (green)
       and $\phi_{23}/\pi$ (blue) in the $SU(3)$ model. Phases $\phi_{13}$
       and $\phi_{23}$ are undefined where $|W_{13}|$ and $|W_{23}|$ are
       zero.}
     \label{fig:AGB_fig_2}
   \end{center}
 \end{figure}

\begin{figure}[!ht]
  \begin{center}
    \includegraphics[width=3.65in,height=5.5in, angle=270]{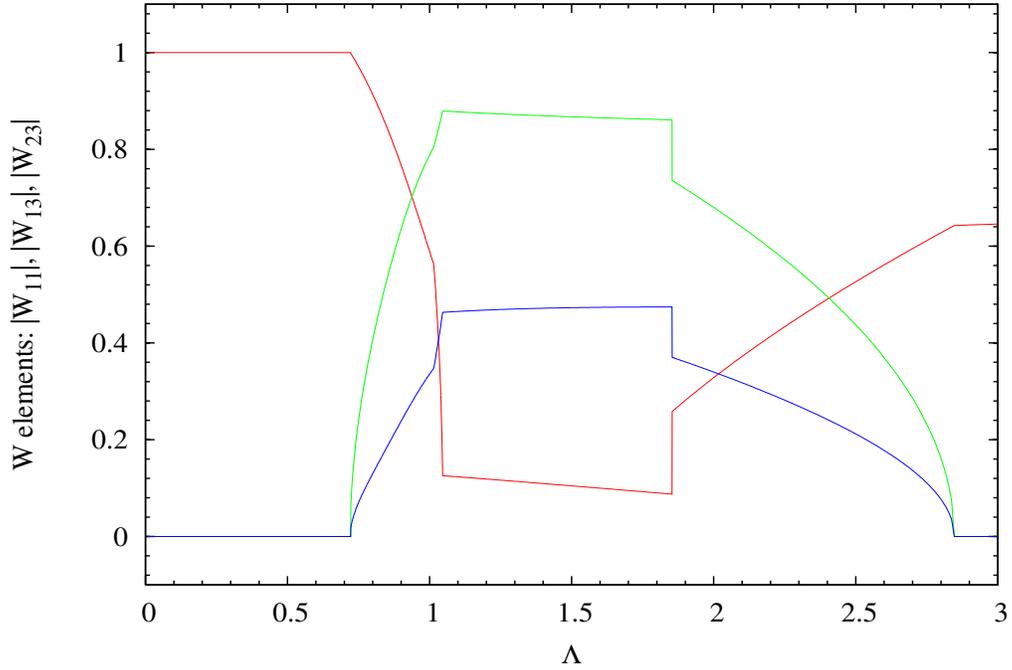}
    \caption{The $W$-magnitudes $|W_{11}|$ (red), $|W_{13}|$ (green) and
     $|W_{23}|$ (blue) in the $SU(3)$ model.}
    \label{fig:AGB_fig_3}
  \end{center}
\end{figure}

\begin{figure}[t]
  \begin{center}
    \includegraphics[width=3.75in,height=5.5in, angle=270]{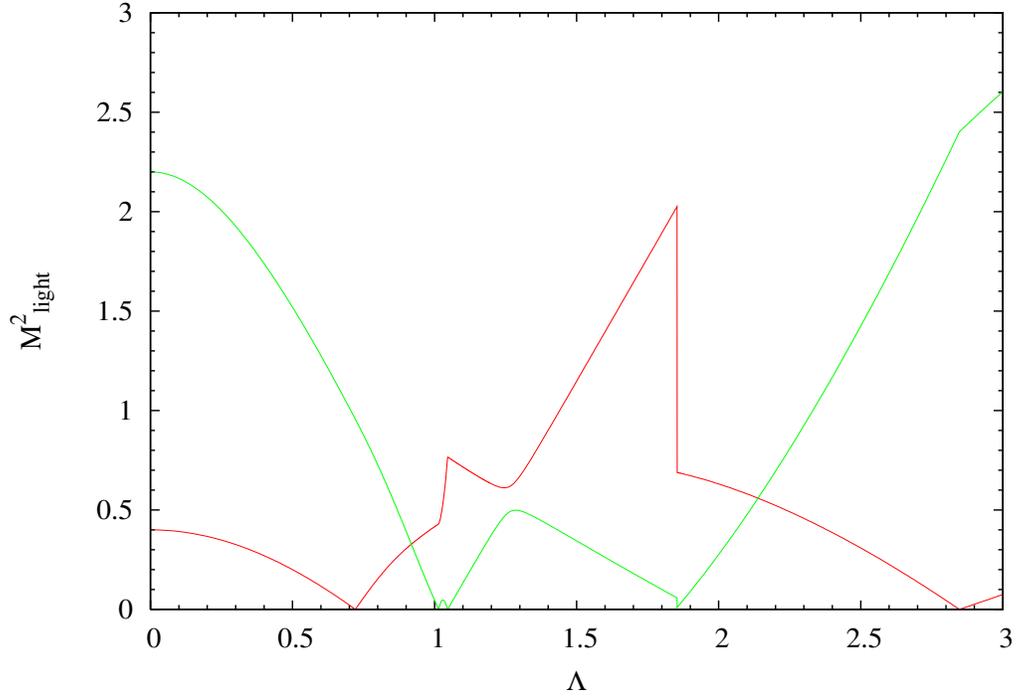}
    \caption{The $M^2$ of the lightest two pseudoGoldstone bosons in the
      $SU(3)$ model.}
    \label{fig:AGB_fig_4}
  \end{center}
\end{figure}

\begin{figure}[!ht]
  \begin{center}
    \includegraphics[width=3.75in,height=5.5in, angle=270]{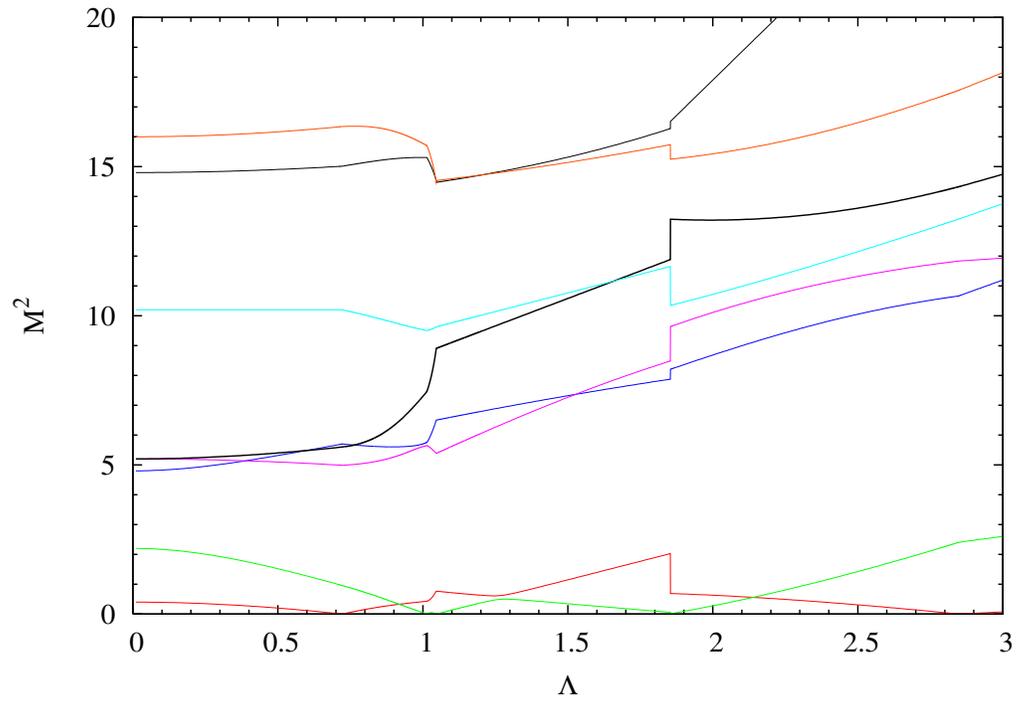}
    \caption{The $M^2$ of all eight pseudoGoldstone bosons in the
      $SU(3)$ model.}
    \label{fig:AGB_fig_5}
  \end{center}
\end{figure}

Vacuum alignment was carried out numerically. For $\Lambda=0$, an initial
guess is made for the phases and angles in $D_{L,R}$ and $K$, and these are
varied to search for a minimum.\footnote{We have not systematically
  established that we have found global minima, but searches with widely
  different inputs have not produced deeper ones.} When an aligning matrix
$W$ is found that minimizes $E$, it is used to calculate the rotated
Hamiltonian $\CH'(W)$ in Eq.~(\ref{eq:HprimeW}) and the PGB matrix $F_\pi^2
M^2_{ab}$ in Eq.~(\ref{eq:Msq}). The eigenvalues and eigenvectors of this
matrix are then determined. Then, $\Lambda$ is increased slightly, the phases
and angles of the $W$ just obtained are used as new inputs, and the procedure
is repeated.  This works well everywhere except at the discontinuous
transition occurring near $\Lambda = 1.9$. Following the mass eigenstates
through that transition is a matter of some judgement --- but not much
import. The results are shown in
Figs.~\ref{fig:AGB_fig_1}--\ref{fig:AGB_fig_5}. There we display the
variation of the minimized vacuum energy, $E(W)$, the phases and magnitudes
of $W_{11}$, $W_{13}$ and $W_{23}$ (these contain phases unlinked to each
other), and the masses of the two lightest PGBs alone and then compared to
the model's other six PGBs.

\begin{figure}[t]
  \begin{center}
    \includegraphics[width=3.75in,height=5.5in, angle=270]{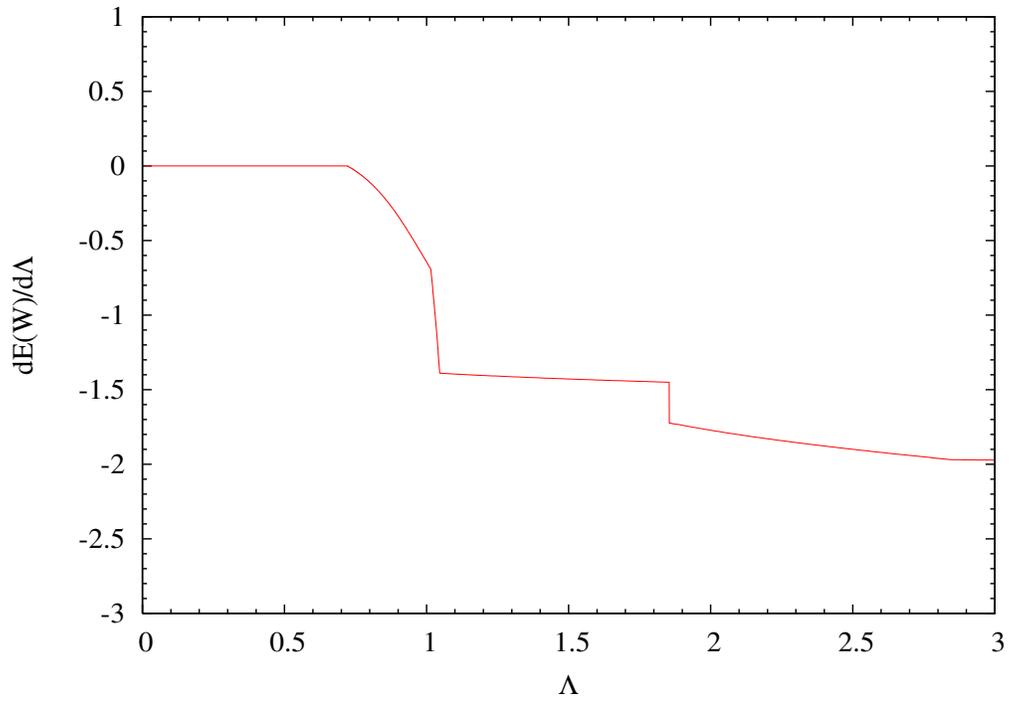}
    \caption{$dE(W)/d\Lambda$ in the $SU(3)$ model.}
    \label{fig:AGB_fig_6}
  \end{center}
\end{figure}

The energy is constant and $W=Z_3 \cdot {\bs 1} $ from $\Lambda = 0$ to
0.7215; this is a CPC phase.\footnote{Recall that $W$ is defined only up to a
  power of $Z_3$.} At this point, there is a transition to a PCP phase in
which $W$ becomes nondiagonal; $\phi_{11}$ still equals $2\pi/3$ but
$\phi_{13}=\pi/6$ and $\phi_{23}=-5\pi/6$. The phases in $\CH'(W)$ are
0,$\pi$ and $\pi/2$. The lightest PGB's $M^2$ goes to zero, and starts to
increase surpassing that of the second lightest PGB near $\Lambda = 0.9$.
That PGB's $M^2$ vanishes at $\Lambda = 1.0140$, then rises and quickly falls
back to zero at $\Lambda = 1.0462$. This small region is a CPV phase with
irrational phases. The region from $\Lambda = 1.0462$ to~1.854 is a CPC phase
with all phases equal $0$ (mod~$\pi$). Up to this point, the energy,
$\phi_{ij}$, $|W_{ij}|$ and all $M^2$ have varied continuously, although
there are obvious discontinuities in the slopes of all but
$E(W)$.\footnote{The phases $\phi_{13}$ and $\phi_{23}$ are not defined below
  $\Lambda = 0.72$ and above 2.85, so their behavior there is not
  discontinuous.}  Here, there is a jump in these quantities and, as can be
seen in Fig.~\ref{fig:AGB_fig_1}, in the slope of $E$. To see it better, we
plot $dE(W)/d\Lambda$ in Fig.~\ref{fig:AGB_fig_6}. This transition is from
the CPC phase to a PCP one.  The lightest PGB appears to become massless, but
it is difficult to tell numerically because of the discontinuous change from
one set of vacua to the another. Finally, there is another transition back to
a CPC phase near $\Lambda = 2.85$. There, $\Lambda_{1222}$ is so large that
$W$ becomes block-diagonal with the mixing elements $W_{13}$ and $W_{23}$
vanishing.

\begin{figure}[!ht]
  \begin{center}
    \includegraphics[width=3.75in,height=5.5in, angle=270]{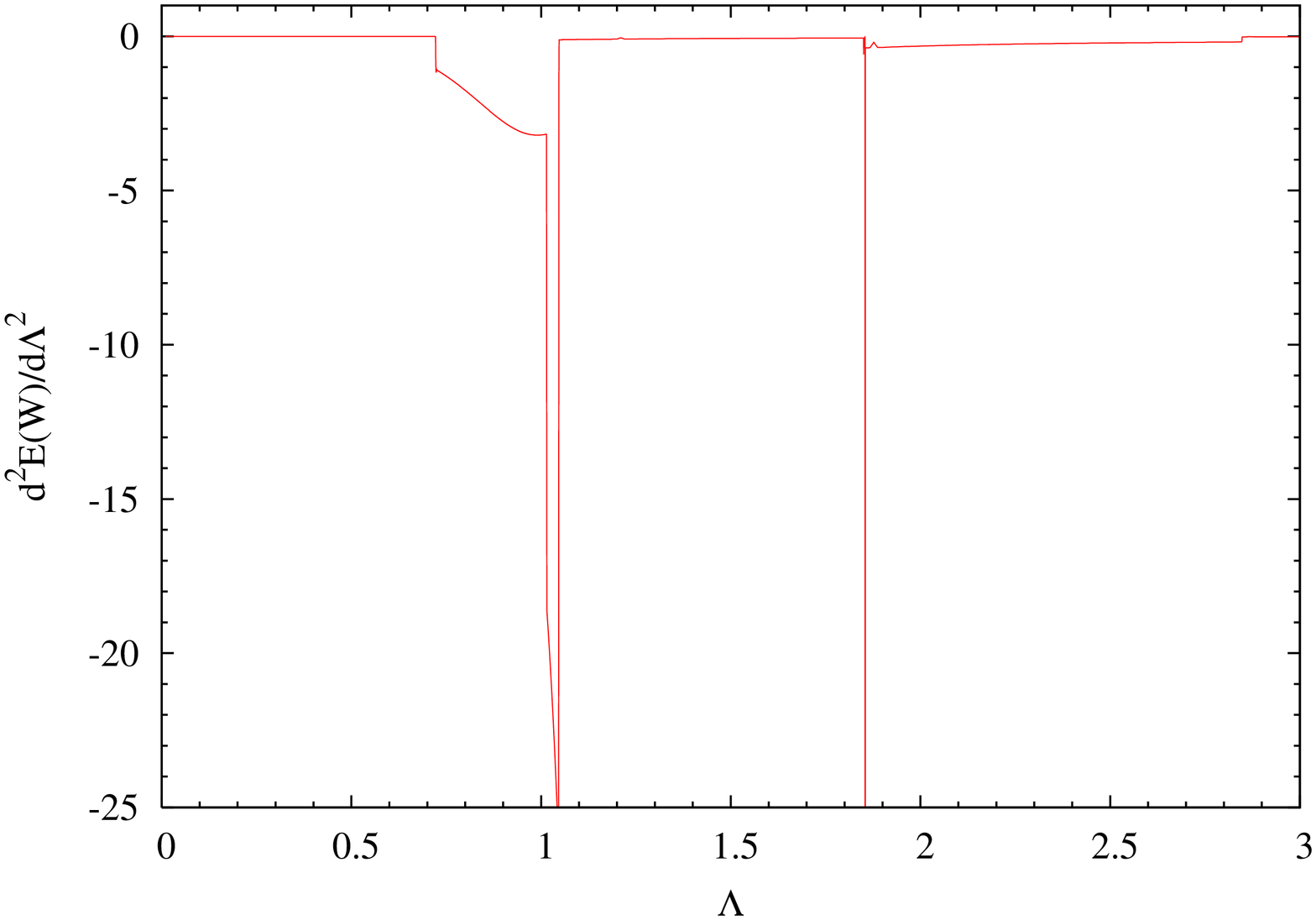}
    \caption{$\dtwoE$ in the $SU(3)$ model.}
    \label{fig:AGB_fig_7}
  \end{center}
\end{figure}
%  
%
%\vfil\eject

We classify the transitions between different CP phases as being of first
order (1-OPT) or second order (2-OPT) depending on whether $dE(W)/d\Lambda$
or $\dtwoE$ is discontinuous at the transition. The second derivative is
plotted in Fig.~\ref{fig:AGB_fig_7}; we will discuss it in the next section.
First-order transitions involve discontinuous changes in $W$-matrix elements.
They occur only at CPC--PCP transitions. The elements of $W$ are continuous
at second-order transitions. They occur at the boundaries between CPC or PCP
regions and CPV ones, or at CPC--PCP boundaries such as $\Lambda = 0.72$
and~2.85 where elements of $W$ continuously become nonzero or vanish.

We stress that the vanishing of an $M^2$ eigenvalue at a phase transition is
not a consequence of increased chiral symmetry; the current corresponding to
the massless boson is still not conserved at the transition.  Rather, the
boson's masslessness is associated with a change in the discrete CP symmetry.
We refer to the two chronically light PGBs of this model as accidental
Goldstone bosons. They remain light because --- in this model and others we
have looked at --- one is never very far from a phase transition.  We explain
in Sec.~III why there are two AGBs in this model.

It is easy to understand why one PGB's $M^2 \ra 0$ at a 2-OPT, $\Lambda =
\Lambda_*$. As $\Lambda < \Lambda_*$ is increased, the true vacuum
corresponding to one CP phase is becoming less stable, while the false vacuum
corresponding to a different phase is becoming more stable. In this false
vacuum, one PGB has $M^2 < 0$.\footnote{There cannot be more than one. In a
  true vacuum, all $M^2 \ge 0$, and it seems most unlikely that two PGB
  masses will vanish at the same $\Lambda_*$ on their way from negative to
  positive values.} In the true vacuum this PGB's positive $M^2$ is
decreasing while it is increasing in the false one. Since the 2-OPT is
continuous, the two $M^2$ trajectories must cross at $M^2 = 0$.  For a 1-OPT,
there is a discontinuous jump in the lightest-$M^2$ as there is for all the
others. Hence, there seems to be no argument for $M^2 \ra 0$.  Nevertheless,
in our calculations for this and other models, the lightest AGB mass appears
to approach zero on one side of the 1-OPT as well. It is obvious that there
are surfaces in the space of the $\Lambda_{ijkl}$ that separate the different
CP phases and, at least for 2-OPT surfaces, an AGB mass vanishes
there.\footnote{We suspect that the order of the phase transition does not
  change as long as new $\Lambda$'s are not introduced. We also note that
  adding new $\Lambda$'s can change the character of a phase, e.g., from PCP
  to CPV if too many phases are linked to be consistent with unitarity.}

There is a clear level-crossing phenomenon in Fig.~\ref{fig:AGB_fig_4}, in
the CPC region near $\Lambda = 1.25$. There we see the two lightest PGBs'
masses approach other and repel.\footnote{The two levels cross, but without
  interaction, in PCP regions, near $\Lambda = 0.9$ and 2.15.} The effect of
this will be seen on the vevs of these states, discussed in Sec.~V.

A comment on the units used for $M^2$ in Figs.~\ref{fig:AGB_fig_4} and
\ref{fig:AGB_fig_5} is in order: The quantity being plotted in these figures
is actually $F_\pi^2 M^2$. In our numerical calculations, we set $\Delta_{TT}
= 1$ so that $\Lambda_{ijkl} \Delta_{TT} = \CO(1)$. But, up to an anomalous
dimension factor for the four-fermion condensate, $\Lambda_{ijkl} \Delta_{TT}
\simeq \Lambda_{ijkl} \Lambda_T^2 F_\pi^4$ where $\Lambda_T \simeq 4\pi
F_\pi$. If, for example, $\Lambda_{ijkl} = (10\Lambda_T)^{-2}$, then the
vertical scale in Figs.~\ref{fig:AGB_fig_4},\ref{fig:AGB_fig_5} is in units
of $10^{-2} F_\pi^4$. The AGB masses are then $M \simle 0.1 F_\pi \simeq
10^{-2} \Lambda_T$.

Finally, we do not believe that these phase transitions and the associated
vanishing of a PGB mass are mere artifacts of our using lowest-order chiral
perturbation theory. Higher-order corrections may shift the surfaces in
$\Lambda$-space separating the phases (not to mention expanding the
dimensions of the space), and they may even eliminate existing transitions or
add new ones. But we see no reason that phase linking, the transitions
between various rational and irrational phase solutions, and the associated
massless states would not occur for $\CH'$ with higher dimensional than
four-fermion operators and vacuum energies involving higher powers of $W$ and
$W^\dag$.

\section*{III. Understanding the Phase Transitions I: \\ The Formula for
  $\dtwoE$}

Considerable insight into the AGBs --- their number and the connection
between their vanishing masses and the behavior of the $W$-phases --- can be
gained from studying $d^2 E(W)/d\Lambda^2$. For definiteness, we continue to
consider a theory in which chiral flavor symmetry $G_f = SU(N)_L \otimes
SU(N)_R$ is spontaneously broken in the vacuum $\rvac$ to $S_f =
SU(N)_V$.\footnote{This discussion and Eq.~(\ref{eq:dtwoEz}) apply to any
  symmetry groups $G_f$ and $S_f$.} The chiral symmetry $G_f$ is also
explicitly broken by an interaction $\CH'$ as in Eq.~(\ref{eq:Hprime0}), for
example. Suppose that $\CH'$ depends {\em linearly} on a parameter $\Lambda$.
Write the vacuum energy of the properly aligned Hamiltonian as\footnote{The
  reason $W=e^{2i{\bs t}\cdot {\bs \omega}}$ is that, for our model's
  symmetry groups, $W = W_L W_R^\dag$ with $W_L = W_R^\dag = e^{i{\bs t}\cdot
    {\bs \omega}}$.}
\be\label{eq:vacE0}
E(W \equiv e^{2i{\bs t}\cdot {\bs \omega}}) =
\lvac \CH'({\bs \omega}) \rvac \equiv \lvac e^{i{\bs Q_5}\cdot {\bs \omega}}
\, \CH' e^{-i{\bs Q_5}\cdot {\bs \omega}} \rvac\,,
\ee
where $\omega_a$, $a=1,\dots,N^2-1$, is a $W$-phase at the minimum. Then (sum
on repeated indices)
\be\label{eq:dtwoEz}
\frac{d^2 E(W)}{d\Lambda^2} = -\CG_{ac}(-{\bs  \omega}) \,
(F_\pi^2 M^2)_{cd} \, \CG_{db}({\bs \omega})
\frac{d\omega_a}{d\Lambda} \frac{d\omega_b}{d\Lambda} 
\equiv -\frac{\partial^2 E(W)}{\partial \omega_a \partial \omega_b} \,
\frac{d\omega_a}{d\Lambda} \frac{d\omega_b}{d\Lambda} \,.
\ee
Equation~(\ref{eq:dtwoEz}) is derived in Appendix~A. Here, $M^2$ is the PGB
squared-mass matrix and $\CG({\bs \omega})$ is the matrix
\be\label{eq:Gmatrix}
\CG_{ab}({\bs \omega}) = \CG_{ba}(-{\bs \omega}) =
\left(\frac{e^{i{\bs F}\cdot{\bs \omega}} - 1}{i{\bs F}\cdot{\bs
          \omega}}\right)_{ab}
      = \sum_{n=0}^\infty \frac{((i{\bs F}\cdot{\bs \omega})^n)_{ab}}
      {(n+1)!}\,,
\ee
and $(F_a)_{bc} = -if_{abc}$ is the adjoint representation of $G_f$. At a
minimum, $\partial^2 E(W)/\partial \omega_a \partial \omega_b$ is a
positive-semidefinite matrix, so that $\dtwoE \le 0$, as seen in
Fig.~\ref{fig:AGB_fig_7}.

To go further with Eq.~(\ref{eq:dtwoEz}), it is convenient to replace $W$ by
its diagonalized form:
\be\label{eq:Wdiag}
 W = e^{2i{\bs t}\cdot {\bs \omega}}
      =  U W_D U^\dag \equiv U \left(e^{2i{\bs t_D}\cdot {\bs
            \omega_D}}\right) U^\dag \,.
 \ee
 Here, $U$ is the $SU(N)$ matrix which diagonalizes $W$ to $W_D$ and ${\bs t}
 \cdot {\bs \omega}$ to ${\bs t_D}\cdot {\bs \omega_D}$.  There are $N-1$
 diagonal phases $\omega_{Da}$, $a = n^2-1$ with $n=2,\dots,N$. They depend
 in complicated ways on the $N^2-1$ phases $\omega_a$ and the parameters in
 $U$. For $t_a \in {\bs N}$, define the real orthogonal matrix $S$ by $S_{ab}
 = 2{\rm Tr}(U^\dag t_a U t_b)$. Then, $\sum_{a=3}^{N^2-1} t_{Da}\omega_{Da}
 \equiv U^\dag (\sum_b t_b \omega_b) U = \sum_{b,c} S_{bc} t_c \omega_b$
 implies
\be\label{eq:omegaD}
\sum_b S_{ba} \omega_b = \left\{\begin{array}{l}
\omega_{Da}\,\,\, {\rm for}\,\, a = 3,8,\dots,N^2-1\\
0\,\,\, {\rm otherwise.}\\
\end{array}\right.
\ee
Next, define $M^2_U$ by
\be\label{eq:MsqU}
(F_\pi^2 M^2)_{cd} = S_{ce} \left(F_\pi^2 M^2_U\right)_{ef}
S^{-1}_{fd}\,.
\ee
For $\CH'$ of the form in Eq.~(\ref{eq:Hprime0}), $M^2_U$ is given by
\bea\label{eq:MsqUform}
\left(F_\pi^2 M^2_U\right)_{ef} &=&
2 \sum_{ijkl} \Lambda^U_{ijkl} \Biggl[
\left(\left\{t_e, t_f\right\}W_D^\dag\right)_{li} \ts W_{Djk} +
\left(W_D\left\{t_e, t_f\right\}\right)_{jk} \ts W_{Dli}^\dag
\nn\\
& &\qquad -2 \left(t_e W_D^\dag\right)_{li} \ts
\left(W_D t_f \right)_{jk} 
-2 \left(t_f W_D^\dag\right)_{li} \ts
\left(W_D t_e\right)_{jk}\ts \Biggr]\Delta_{TT} \,;\\
\Lambda^U_{ijkl} &=& \sum_{i'j'k'l'} \Lambda_{i'j'k'l'}U^\dag_{ii'} U_{j'j}
U^\dag_{kk'} U_{l'l}\,.
\eea
The relation $f_{def} S_{da} S_{eb} S_{fc} = f_{abc}$ and
Eq.~(\ref{eq:omegaD}) imply $S^{-1}{\bs F}\cdot {\bs \omega} S = {\bs
  F_D}\cdot {\bs \omega_D}$, where $F_{Da}$ is a diagonal generator in the
adjoint representation. Thus, Eq.~(\ref{eq:dtwoEz}) can be cast in the form
\be\label{eq:dtwoEzdiag}
\frac{d^2 E(W)}{d\Lambda^2} = -\left(\CG(-{\bs
  \omega_D})\, F_\pi^2 M^2_U \, \CG({\bs \omega_D})\right)_{ab}
\left[\frac{d\omega_{Da}}{d\Lambda} - \omega_c
  \frac{dS^{-1}_{ac}}{d\Lambda}\right] 
\left[\frac{d\omega_{Db}}{d\Lambda} - \omega_d
  \frac{dS^{-1}_{bd}}{d\Lambda}\right] \,.
\ee
This is our key equation.

\begin{figure}[t]
  \begin{center}
    \includegraphics[width=3.75in,height=5.5in, angle=270]{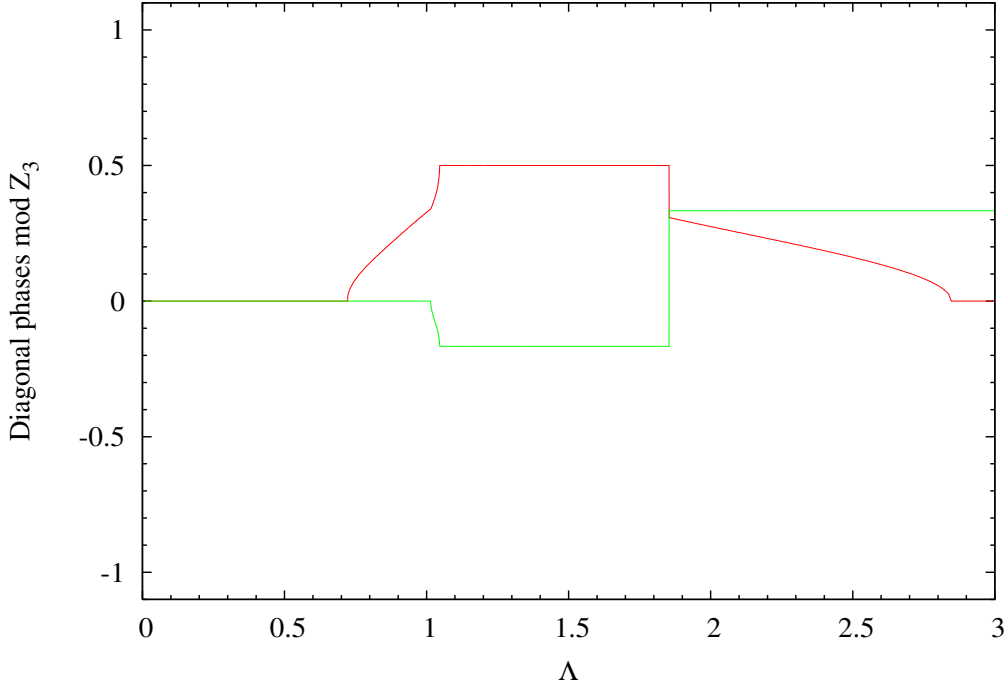}
    \caption{The normalized diagonal phases $\widehat\omega_{D3}/\pi$ (red) and
      $\widehat\omega_{D8}/\pi$ (green) in the $SU(3)$ model.}
    \label{fig:AGB_fig_8}
  \end{center}
\end{figure}
\begin{figure}[t]
  \begin{center}
    \includegraphics[width=3.65in,height=5.5in, angle=270]{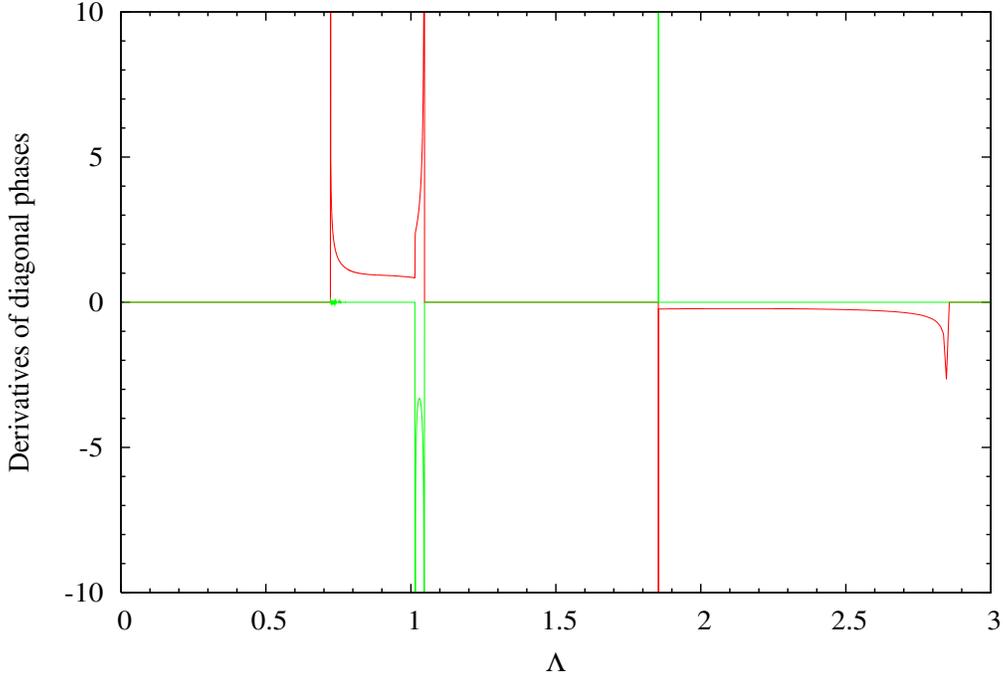}
    \caption{$d(\widehat\omega_{D3}/\pi)/d\Lambda$ (red) and
      $d(\widehat\omega_{D8}/\pi)/d\Lambda$ (green) in the $SU(3)$ model.}
    \label{fig:AGB_fig_9}
  \end{center}
\end{figure}
\begin{figure}[!ht]
  \begin{center}
    \includegraphics[width=3.65in,height=5.5in, angle=270]{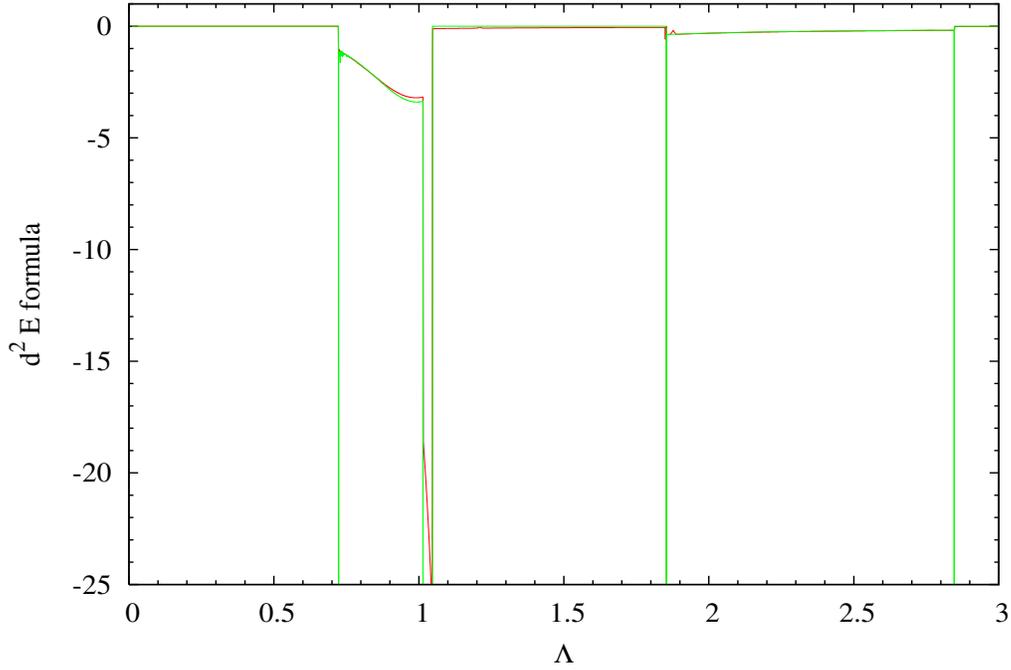}
    \caption{A comparison of $\dtwoE$ (red) and the right-hand
      side of Eq.~(\ref{eq:dtwoEzapprox}) (green) in the $SU(3)$ model.}
    \label{fig:AGB_fig_10}
  \end{center}
\end{figure}

In Sec.~II we saw that all $W$-phases $\phi_{ij}$ are rational multiples of
$\pi$ in the CPC and PCP phases. In Fig.~\ref{fig:AGB_fig_8} we plot the
normalized diagonal phases $\homega = \omega_{Da}/\sqrt{n(n-1)/2}$ for the
$SU(3)$ model ($n=2,3$). We see that in CPC phases, both $\homega$ are
rational multiples of $\pi$; in PCP phases, only $\widehat\omega_{D8}$ is
rational; in the CPV phase, both are irrational. This will be explained in
Sec.~IV. It is remarkable that, even though $\widehat\omega_{D3}$ is
irrational in the PCP phases, all $\phi_{ij}$ are rational
there.\footnote{The definition of the $\homega$ is convention-dependent. The
  scheme we use for calculating the $\homega$ is this: Starting at the
  initial $\Lambda$, here zero, the matrix $W$ is diagonalized and the phases
  of its eigenvalues --- its eigenphases $\eta_i$ --- are determined. A
  multiple of $2\pi/N$ is subtracted from them so that $\sum_{i=1}^N \eta_i =
  0$.  The eigenvalues are then ordered so that ${\rm Re}(e^{i\eta_i}) \le
  {\rm Re}(e^{i\eta_{i+1}})$. Then, $\widehat \omega_{D,N^2-1} =
  -\eta_N/(N-1)$, $\widehat \omega_{D,(N-1)^2-1} = -(\eta_{N-1} +
  \eta_N/(N-1))/(N-2)$, etc.  As $\Lambda$ is increased, the procedure is
  repeated, requiring the changes in the $\eta_i$ and the $\homega$ to be
  continuous, except at a 1-OPT. If necessary, the multiple of $2\pi/N$
  subtracted from the $\eta_i$ is changed to keep their evolution continuous.
  These subtraction changes typically occur at 2-OPTs. The discontinuous
  changes at a 1-OPT are also kept as small as possible. In the CPC and PCP
  phases, nonzero $\homega$ are actually rational multiples of $\pi$ to about
  a part in $10^3$, whereas the $\phi_{ij}$ are rational to computer
  accuracy. As we discuss in Sec.~IV, the near rationality of the $\homega$
  appears to be an unintentional artifact of the way we chose the
  $\Lambda_{ijkl}$.}

One sees in Fig.~\ref{fig:AGB_fig_8} that the slope of one or both of the
$\homega$ is singular at every 2-OPT ($d\widehat\omega_{D3}/d\Lambda$ is
merely discontinuous at $\Lambda=1.0140$) while both $\homega$ are
discontinuous at the 1-OPT at $\Lambda = 1.85$. Looking back at
Fig.~\ref{fig:AGB_fig_2}, this behavior is clearly reflected in all the
$\phi_{ij}$; it is especially dramatic at the 2-OPTs near $\Lambda = 1$. The
slopes $d\homega/d\Lambda$ are plotted in Fig.~\ref{fig:AGB_fig_9}. Away from
the phase transitions, they are not large except in the narrow CPV phase
where the $\homega$ are rapidly varying.\footnote{We have numerically studied
  an $SU(4)$ model and found very similar features to the ones described
  here. One difference is that the CPV phase in that model is wider. This is
  not important; in fact, it is surprising that the CPV phase in the $SU(3)$
  model is so narrow.} The singular behavior of the $\homega$ in
Fig.~\ref{fig:AGB_fig_8} is just what we expect of order parameters at first
and second-order phase transitions. Therefore, we interpret the diagonal
phases $\omega_{Da}$ as the order parameters for the phase transitions we've
been observing. Here, however, the transitions are between different phases
of a {\em discrete} symmetry.

In general, the $dS^{-1}_{ac}/d\Lambda$ in Eq.~(\ref{eq:dtwoEzdiag}) are
small. Thus, $\dtwoE$ is well approximated by keeping only the
$(d\omega_{Da}/d\Lambda) \, (d\omega_{Db}/d\Lambda)$ terms in
Eq.~(\ref{eq:dtwoEzdiag}). Just how good this approximation is can be seen by
looking at the region $\Lambda = 1.05$ to~1.85 in Fig.~\ref{fig:AGB_fig_7}.
There $d\omega_{Da}/d\Lambda \equiv 0$, while $\dtwoE$ is negative, but very
small. If we drop the $dS^{-1}_{ac}/d\Lambda$ terms,
Eq.~(\ref{eq:dtwoEzdiag}) simplifies greatly because $(F_{Dc})_{ab} = 0$ and
$\CG_{ab}({\bs \omega_D}) = \delta_{ab}$ when index $a$ or $b =
3,8,\dots,N^2-1$. Then
\be\label{eq:dtwoEzapprox}
\frac{d^2 E(W)}{d\Lambda^2} \cong -\left(F_\pi^2 M^2_U\right)_{ab}
\frac{d\omega_{Da}}{d\Lambda} \, \frac{d\omega_{Db}}{d\Lambda}\,.
\ee
In Fig.~\ref{fig:AGB_fig_10} we compare $\dtwoE$ with the right-hand side of
Eq.~(\ref{eq:dtwoEzapprox}). The agreement is excellent except in the narrow
CPV region with rapidly varying phases. There, the discrepancy is due both to
the neglect of the $\omega_c dS^{-1}_{ac}/d\Lambda$-terms and the difficulty
of computing the derivatives as they become divergent.

Equation~(\ref{eq:dtwoEzapprox}) makes a clear connection between the
lightest PGBs, the ones we call AGBs, and the diagonal phases $\omega_{Da}$.
We believe the association is one-to-one, and that is why the $SU(3)$ model
has two AGBs.\footnote{We have examined larger $SU(N)$ models and never found
  more than $N-1$ especially light PGBs. Of course, this one-to-one
  connection is applicable only so long as all $\pi_a$ symmetries are
  explicitly broken so that there are no true Goldstone bosons.} At 2-OPTs,
the $\omega_{Da}$ are continuous, but at least some $d\omega_{Da}/d\Lambda$
are divergent.  Meanwhile, $\dtwoE$ is finite, though discontinuous. This is
possible only if a zero eigenvalue of the PGB $M^2$-matrix appears exactly at
the transition to cancel singularities in the
$d\omega_{Da}/d\Lambda$.\footnote{An analytic example is given for Dashen's
  $SU(3)$ model in Appendix~B.} This is another reason we believe that the
vanishing of AGB masses at phase transitions is not an artifact of
lowest-order chiral perturbation theory. At a 1-OPT at $\Lambda_*$,
$\omega_{Da}$ is discontinuous and $d\omega_{Da}/d\Lambda \propto
\delta(\Lambda - \Lambda_*)$. On the other hand, all the PGB masses are
discontinuous there, so we expect $\dtwoE \propto \delta(\Lambda -
\Lambda_*)$, i.e., a discontinuous slope in $E(W)$, as well.

\section*{IV. Understanding the Phase Transitions II: \\ The Character of
  $\widehat W = D_R D_L K$}

In Sec.~II we showed that, in a basis in which the aligning matrix is $\whW =
D_R W D_R^\dag = D_R D_L K$, the LR terms in $\CH'(\whW)$ are real in the PCP
and CPC phases. The matrix $\whW$ has the same eigenvalues as $W$, therefore
the same diagonal phases $\homega$. However, it is easier to analyze the
possibilities for the $\homega$ by considering $\whW$.

Consider first the CPC phase. In that case, $\whW = e^{2im\pi/N} \wtW$ where
$\wtW$ is an $SO(N)$ matrix and $m=0,\dots,N-1$. Denote $\wtW$'s
eigenvalues by $e^{i\eta_i}$, $i = 1,\dots,N$ where the eigenphases satisfy
$\sum_i\eta_i = 0$ (mod $2\pi$). If $N$ is even, the eigenphases form
conjugate pairs, $(e^{i\eta_i}, e^{-i\eta_i})$ for $i=1,\dots,N/2$. If $N$ is
odd, one eigenvalue, say $e^{i\eta_N}$, is $+1$. The ordering of the
$\eta_i$ is arbitrary. Given an ordering, we can calculate the diagonal
phases from
\bea\label{eq:diagphases}
\widehat \omega_{D,N^2-1} &=& -\left(\frac{1}{N-1}\right)(2m\pi/N +
  \eta_N)\,, \nn\\ 
\widehat \omega_{D,(N-1)^2-1} &=& -\left(\frac{1}{N-2}\right)
((2m\pi + \eta_N)/(N-1) + \eta_{N-1})\,,\dots 
\eea
Because of the $Z_N$ ambiguity in $\wtW$, we can set $m=0$ if we wish.

Now, {\em if} $\wtW$ is also symmetric, then all its eigenvalues are real,
therefore equal $\pm 1$, with an even number of $-1$'s. All its eigenphases
of $\whW$ would be rational multiples of $\pi$ and, then, so would their
linear combinations forming the $\homega$.  In the models we studied
numerically, $\wtW$ is symmetric to about a part in $10^3$ in all nontrivial
CPC phases, i.e., when $\wtW$ is not merely proportional to the identity.
Hence, the $\homega$ are rational to about the same accuracy in these
calculations. The difference from exactly rational phases is not visible in
the CPC regions of Fig.~\ref{fig:AGB_fig_8}. This closeness to rational
phases is tantalizing, but we believe it is an unintended artifact of the way
we chose the couplings $\Lambda_{ijkl}$ in the $SU(3)$ model. Those couplings
seem to favor minimizing $E$ with a symmetric $\wtW$; we have modified them
to make $\wtW$ non-symmetric in a CPC phase.

Turning to the PCP case, in which the phases $\phi_{ij}$ of $W_{ij}$ are
different rational multiples of $\pi$, we have identified two subphases:
PCP-1 in which $\whW = e^{i\phi} \wtW$ with $\wtW$ a real $O(N)$ matrix, and
PCP-2 in which $\whW$ cannot be written in this way. In PCP-1, which is what
we observed in our $SU(3)$ model, $\phi = 2m\pi/N$ if $\det \wtW = 1$, while
$\phi = (2m+1)\pi/N$ if $\det \wtW = -1$. If $N$ is odd and $\det\wtW = -1$,
we can change the sign of $\wtW$ and take $\phi = 2m\pi/N$. For odd $N$,
then, the eigenvalues of $\wtW$ form $(N-1)/2$ pairs, $(e^{i\eta_i},
e^{-i\eta_i})$ plus one real eigenvalue, $e^{i\eta_N} = 1$ and, so, $\wtW$
has $2n+1$ truly rational eigenphases, $n=0,1,\dots, (N-1)/2$. As in
Eq.~(\ref{eq:diagphases}), we can define $\widehat \omega_{D,N^2-1} =
-2m\pi/N(N-1)$. If $N$ is even and $\det \wtW = 1$, $\wtW$ has $2n =
0,2,\dots,N$ rational phases. In this case, there may be no rational
$\homega$ even though all $W$-phases $\phi_{ij}$ are rational. If $\det\wtW =
-1$, there must be a real pair of eigenphases, $(1,-1)$, so there are will be
at least two rational $\homega$. We can choose them to be $\widehat
\omega_{D,N^2-1} = -((2m+1)\pi/N \pm \pi)/(N-1)$ and $\widehat
\omega_{D,(N-1)^2-1} = -((2m+1)\pi \pm \pi)/(N-1)(N-2)$.

Finally, in a PCP-2 phase, there is no argument that any of the $\homega$
are rational. The same is of course true in a CPV phase, and we find only
irrational phases in both.

\section*{V. VEVs of the AGBs}

In this section we investigate whether AGBs can serve as light composite
Higgs bosons. We have seen that they are usually much lighter than the scale
$\Lambda_T \simeq 4\pi F_\pi$ of their strong binding interaction.  Having
associated the AGBs with the diagonal phases $\omega_{Da}$ and, in turn,
identified these as the order parameters of the various CP phases, it is
natural to connect the vacuum expectation values of the AGBs with these
phases. The question studied here is whether these vevs can also be much less
than $\Lambda_T$.

In a nonlinear sigma-model formulation of the $G_f = SU(N)_L \otimes SU(N)_R$
model, we would replace $\ol T_{Rj} T_{Li}$ by $F_\pi^3 \Sigma_{ij}$, where
$\Sigma = \exp{(2i{\bs t}\cdot{\bs \pi}/F_\pi)}$.\footnote{This normalization
  of $\Sigma$ guarantees that the axial current $j_{5\mu}^a$ it generates
  creates $\pi_a$ from the vacuum with strength $F_\pi$.} Under a $G_f$
transformation, $\Sigma \ra W_L\Sigma W_R^\dag$. Minimizing the energy $E(W)$
in this formulation amounts to determining the vacuum expectation values
$\langle\pi_a\rangle = \lvac\pi_a\rvac$ in the tree approximation. Thus,
these vevs are related to the minimizing-$W$ phases $\omega_a$ by
\be\label{eq:vevs}
\langle\pi_a\rangle = \omega_a F_\pi\,.
\ee
To determine the vevs of the $N-1$ AGBs of the model, we write
\bea\label{eq:Msqomegasq}
 M^2_{ab}\,\omega_a \omega_b  &=& M^2_{ab} S_{ac} S_{bd}\, \omega_{Dc}
 \omega_{Dd} \nn\\ 
 &=& (M_U^2)_{ab}\, \omega_{Da} \omega_{Db} = (VM_D^2 V^{-1})_{ab}\, \omega_{Da}
 \omega_{Db}\,,
\eea
where $V$ is the $SO(N)$ matrix which diagonalizes $M_U^2 = S^{-1}M^2 S$ to
$M_D^2$. The mass eigenstate vevs $v_a$, in particular, those of the AGBs,
are then
\be\label{eq:AGBvev}
v_a = V_{ba} \,\omega_{Db} F_\pi = (V^{-1}S^{-1})_{ab} \,\omega_b F_\pi\,.
\ee
This definition of the AGB vevs is independent of the convention used to
define the $\omega_{Da}$. Note that, so long as vacuum alignment preserves
electric charge conservation, $W_{ij} = \delta_{ij}$ in electrically charged
sectors and all AGBs are electrically neutral.

\begin{figure}[t]
  \begin{center}
    \includegraphics[width=3.55in,height=5.5in, angle=270]{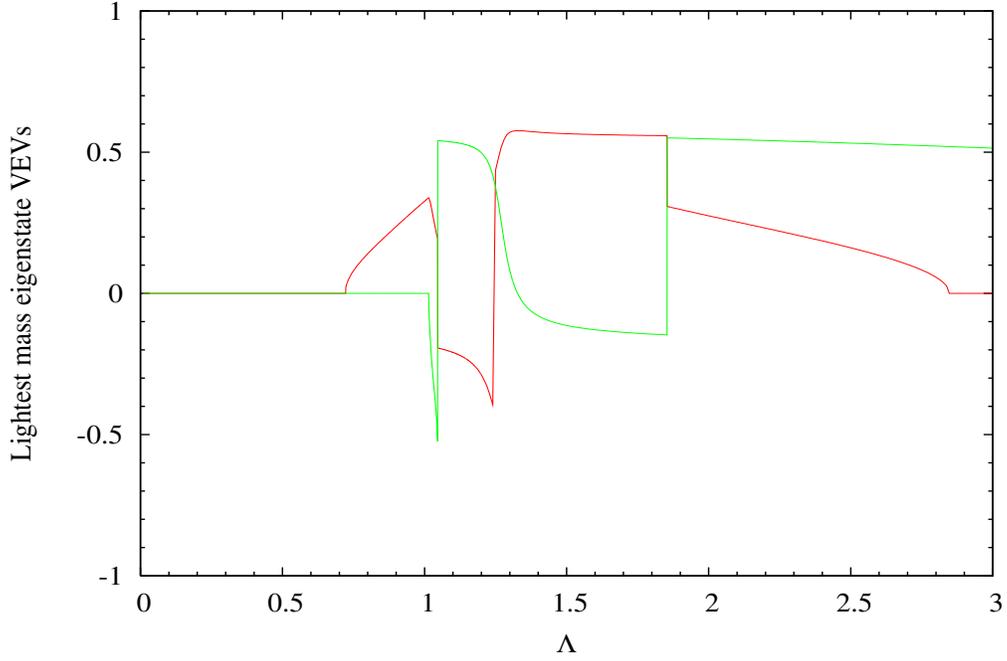}
    \caption{The vevs $(v_a/F_\pi)/\pi$ of the lightest two pseudoGoldstone
      bosons in the $SU(3)$ model. The colors match those in
      Fig.~\ref{fig:AGB_fig_4}.}
    \label{fig:AGB_fig_11}
  \end{center}
\end{figure}
\begin{figure}[!ht]
  \begin{center}
    \includegraphics[width=3.55in,height=5.5in, angle=270]{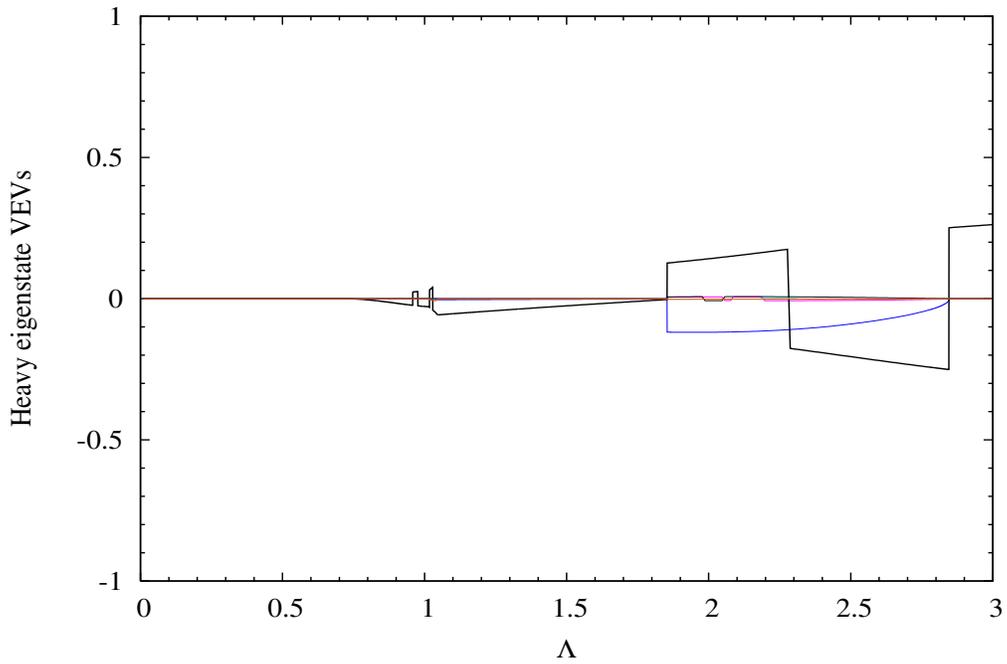}
    \caption{The vevs $(v_a/F_\pi)/\pi$ of the six heavier pseudoGoldstone
      bosons in the $SU(3)$ model. The colors match those in
      Fig.~\ref{fig:AGB_fig_5}.}
    \label{fig:AGB_fig_12}
  \end{center}
\end{figure}

An AGB may be a suitable light composite Higgs if $|v_a/F_\pi| \ll 4\pi$.
These vevs are plotted in Fig.~\ref{fig:AGB_fig_11} for the two lightest AGBs
of the $SU(3)$ model. They are indeed generally small, with $|v_a| \simle
0.03$--$0.1 \Lambda_T$ in all CP phases. Similar results are obtained in an
$SU(4)$-model calculation. The AGBs' vevs tend to track $\omega_{D3}$ and
$\omega_{D8}$, except near $\Lambda = 1.25$. These small vevs seem to be due
to the $1/N$ factors in Eq.~(\ref{eq:diagphases}) and to the fact that
$|V_{ab}| < 1$.  Changes in the vevs due to higher-corrections to $\CH'$
should be small unless those corrections induce a first-order phase
transition. The rapid variation in the vevs near $\Lambda = 1.25$ is due to
the level-crossing visible there in Fig.~\ref{fig:AGB_fig_4}. We have seen
the same phenomenon analytically in the isospin-violating version of the
$SU(3)$ model described in Appendix~B. Finally, if the $G_f$ symmetries were
gauged with coupling~$g$, the AGBs would give masses $\sim gv_a$ to the gauge
bosons that are very much less than the underlying dynamical scale
$\Lambda_T$.

The mass eigenstate vevs of the heavier PGBs are shown in
Fig.~\ref{fig:AGB_fig_12}. They are generally very small, or at most
comparable to those of the light AGBs.\footnote{The sign reversals above
  $\Lambda = 2$ have no physical significance.} This confirms that the light
AGBs correspond to the diagonal phases $\omega_{Da}$. Further, if the heavier
PGBs are coupled to gauge bosons, they generally contribute negligibly to
their mass, and never more than the AGBs do. From an experimental point of
view, one would probably conclude that the gauge symmetries are broken by
light composite Higgs bosons at a scale well below $\Lambda_T$. Small vevs
for the heavier PGBs raise the interesting possibility that a heavy composite
Higgs can naturally give small masses to gauge bosons, with no contribution
coming from the light AGBs.  Experimentally, the only sign of the gauge
symmetry's breaking at energies of order $v_a$ would be a (temporary)
breakdown of perturbative unitarity!

\section*{VI. Summary and Future Work}

In this paper we studied vacuum alignment in theories in which the global
chiral symmetry $G_f$ of a set of $N$ massless Dirac fermions is broken both
spontaneously by their strong interactions and explicitly by terms in a weak
perturbation $\CH'$. This perturbation is chosen to give mass to all the
Goldstone bosons of the spontaneous symmetry breaking. We showed that, as a
coupling parameter $\Lambda$ in $\CH'$ is changed, the system moves through
various phases of the discrete symmetry, CP.  We identified three main
phases: CP-conserving, in which the aligning matrix $W \in SU(N)$ is real up
to a $Z_N$ factor and the aligned Hamiltonian $\CH'(W)$ is real;
CP-violating, in which $W$ and $\CH'(W)$ are essentially complex; and a new
phase, pseudoCP-conserving, in which the phases in $W$ are different rational
multiples of $\pi$ and so are the phases of $\CH'(W)$. For the class of
models we studied, it was actually possible in the PCP phase to make a
transformation that rendered the explicit $G_f$-breaking terms in $\CH'(W)$
real.

Most important, we found that the transitions between different CP phases are
of classic first or second-order, defined as whether the first or second
derivative of the vacuum energy $E(W) = \lvac\CH'(W)\rvac$ with respect to
$\Lambda$ is discontinuous at the transition. At all these transitions a
pseudoGoldstone boson's mass vanishes. Following Dashen~\cite{Dashen:1971et},
we call these accidental Goldstone bosons, AGBs, but we argued that their
presence is not a mere consequence of the lowest-order chiral perturbation
theory we employ to calculate their masses. Rather, they are a necessary
consequence of the CP-phase transitions, phenomena we believe transcend our
$\CO(\CH')$ approximation. The relative frequency of CP-phase
transitions makes AGBs common: there generally seem to be several such
states, much lighter than the other PGBs. We derived a remarkable formula for
$d^2E(W)/d\Lambda^2$ that establishes a one-to-one correspondence between the
AGBs and the eigenphases $\omega_{Da}$ of the diagonalized form $W_D$ of $W$.
In the $SU(N)$ models we studied, $W_D =
\exp{(2i\sum_{a=3}^{N^2-1}t_{Da}\,\omega_{Da})}$ and there are $N-1$ AGBs.
The vanishing of an AGB mass at some $\Lambda = \Lambda_*$ is directly
correlated with the singular behavior of its corresponding combination of
$\omega_{Da}$.

The AGB masses are naturally much less than the scale $\Lambda_T \simeq 4\pi
F_\pi$ of their strong binding interaction. Equally interesting, we found
that their vacuum expectation values also are often much less than
$\Lambda_T$. Thus, they are prototypes for light composite Higgs bosons for
electroweak symmetry breaking. To make a realistic model, we have to find a
way to embed $SU(2) \otimes U(1)$ into the AGBs' symmetry group $G_f$ without
their constituent $T$-fermions' condensates breaking electroweak symmetry at
$\Lambda_T$. One way that does {\em not} work is a technicolor-like scheme
with $N$ doublets, $T_{L,R\,i} = (U,D)_{L,R\, i}$, and a chiral $SU(2N)_L
\otimes SU(2N)_R$ symmetry breaking down to $SU(2N)$. These fermions must
transform vectorially under $SU(2) \otimes U(1)$, with weak hypercharges
$Y_i$. The $Y_i$ must be chosen so that the PGB-mass generating $\CH'$ is
$SU(2) \otimes U(1)$-invariant. Then it is impossible for $\Sigma = e^{2i{\bs
    t}\cdot{\bs \pi}}$ to develop a vacuum expectation value which both
conserves electric charge, $Q=T_3+Y$, and breaks electroweak symmetry in the
correct way; in particular, the $U(1)$ remains unbroken. Another difficult
problem is coupling the AGBs to quarks and leptons so that their vevs can
give them mass. Compounding that difficulty is the need to avoid unwanted
flavor-changing neutral current interactions.  Presumably, one must be in a
PCP phase so that weak, but not strong, CP violation is transmitted to the
quarks through the Yukawa couplings to the
$\Sigma$-field~\cite{Martin:2004ec}.

\section*{Acknowledgements} We have benefited from discussions and previous
work done with Estia Eichten and Tongu\c c Rador. We have also profited from
conversations with Nima Arkani-Hamed, Tom Appelquist, Bill Bardeen, Antonio
Castro-Neto, Claudio Chamon, Martin Schmaltz, Bob Shrock, Witold Skiba, and
Erick Weinberg. KL thanks the Laboratoire d'Annecy-le-Vieux de Physique
Theorique, Annecy, France for its hospitality and partial support for this
research during the summer of 2004.  This research was also supported by the
U.~S.~Department of Energy under Grant~No.~DE--FG02--91ER40676.

\section*{Appendix A: Derivation of the Formula for $\dtwoE$}

Consider a theory in which the chiral flavor symmetry $G_f$ is spontaneously
broken in the vacuum $\rvac$ to $S_f$. The chiral symmetry $G_f$ is also
explicitly broken by an interaction $\CH'$. Suppose that $\CH'$ depends {\em
  linearly} on a parameter $\Lambda$. Write the minimized vacuum energy as
\be\label{eq:vacE0two}
E(W_0 \equiv e^{i{\bs t}\cdot {\bs \omega_0}}) =
\lvac \CH'({\bs \omega_0}) \rvac \equiv \lvac e^{i{\bs Q}\cdot {\bs \omega_0}}
\, \CH' \, e^{-i{\bs Q}\cdot {\bs \omega_0}} \rvac\,.
\ee
Here, $Q_a$ is a generator of $G_f$ and $t_a$ is its matrix representation.
Charges $Q_a \in S_f$ annihilate $\rvac$; charges in $G_f/S_f$ create a
Goldstone boson $\pi_a$ from the vacuum with strength $F_\pi$. We have
reintroduced the subscript ``0'' to emphasize that $W_0$ is the unitary
aligning matrix which minimizes the vacuum energy.

Let us study how the minimized energy changes as we vary $\Lambda$. The
first derivative is (sum on repeated indices)
\be\label{eq:first} \frac{dE(W_0)}{d\Lambda} = \lvac e^{i{\bs Q}\cdot {\bs
    \omega_0}}\, \frac{\partial \CH'}{\partial \Lambda} \, e^{-i{\bs Q}\cdot
  {\bs \omega_0}} \rvac + \left[\frac{\partial}{\partial \omega_{a}}
  \lvac \CH'({\bs \omega}) \rvac \right]_{{\bs \omega} = {\bs \omega_0}}
\frac{d\omega_{a,0}}{d\Lambda} \,.
\ee
The second term vanishes because $E(W)$ is stationary at extrema.
Differentiating again, and using our linearity assumption,
$\partial^2\CH'/\partial \Lambda^2 = 0$, we get
\bea\label{eq:second}
\frac{d^2E(W_0)}{d\Lambda^2} &=& \left[\frac{\partial}{\partial
    \omega_{a}} \lvac e^{i{\bs Q}\cdot {\bs \omega}}\,
\frac{\partial \CH'}{\partial \Lambda}\, e^{-i{\bs Q}\cdot {\bs \omega}}\rvac
\right]_{{\bs \omega} = {\bs \omega_0}} \frac{d\omega_{a,0}}{d\Lambda} \nn\\
&=& i\CG_{ba}({\bs \omega_0})\, \lvac \bigl[Q_b, \, e^{i{\bs Q}\cdot {\bs
    \omega_0}} \, \frac{\partial \CH'}{\partial \Lambda}\, e^{-i{\bs Q}\cdot
    {\bs \omega_0}}\bigr]\rvac \, \frac{d\omega_{a,0}}{d\Lambda}\,,
\eea
where
\be\label{eq:Gmatrixtwo}
\CG_{ab}({\bs \omega}) = \CG_{ba}(-{\bs \omega}) =
\left(\frac{e^{i{\bs F}\cdot{\bs \omega}} - 1}{i{\bs F}\cdot{\bs
          \omega}}\right)_{ab}
      = \sum_{n=0}^\infty \frac{((i{\bs F}\cdot{\bs \omega})^n)_{ab}}
      {(n+1)!}\,,
\ee
and $(F_a)_{bc} = -if_{abc}$ is the adjoint representation of $Q_a$. The
proof of the second equality in Eq.~(\ref{eq:second}) will be given below.
Now,
\bea\label{eq:ID}
\lvac \bigl[Q_b, \, e^{i{\bs Q}\cdot {\bs
    \omega_0}} \, \frac{\partial \CH'}{\partial \Lambda}\, e^{-i{\bs Q}\cdot
    {\bs \omega_0}}\bigr]\rvac &=& \frac{d}{d\Lambda} \lvac \bigl[Q_b, \,
 e^{i{\bs Q}\cdot{\bs \omega_0}} \, \CH'\, 
e^{-i{\bs Q}\cdot{\bs \omega_0}}\bigr]\rvac \\
&& \,\, -\frac{\partial}{\partial \omega_{c,0}}
 \lvac \bigl[Q_b, \,
 e^{i{\bs Q}\cdot{\bs \omega_0}} \, \CH'\, 
e^{-i{\bs Q}\cdot{\bs \omega_0}}\bigr]\rvac \, \frac{d\omega_{c,0}}{d\Lambda}
\nn \,.
\eea
The first term on the right vanishes by the extremal condition on $E(W)$ (see
Eq.~(\ref{eq:extrema}) below). The second term may be rewritten using
Eqs.~(\ref{eq:second}) and~(\ref{eq:Msq}):
\bea\label{eq:IDtwo}
\frac{\partial}{\partial \omega_{c,0}} \lvac \bigl[Q_b, \, e^{i{\bs
    Q}\cdot {\bs \omega_0}} \, \CH'\,
e^{-i{\bs Q}\cdot {\bs \omega_0}}\bigr]\rvac &=& i\CG_{dc}({\bs
  \omega_0})\lvac \bigl[Q_b,\bigl[Q_d,\, \CH'({\bs
  \omega_0})\bigr]\bigr]\rvac \nn\\
&\equiv& -i  \left(F_\pi^2 M^2 \CG({\bs \omega_0})\right)_{bc}\,.
\eea
This gives the desired result:
\be\label{eq:dtwoEztwo}
\frac{d^2 E(W_0)}{d\Lambda^2} = -\left(\CG(-{\bs \omega_0}) \,
F_\pi^2 M^2\, \CG({\bs \omega_0})\right)_{ab}
\frac{d\omega_{a,0}}{d\Lambda} \frac{d\omega_{b,0}}{d\Lambda} \,.
\ee

The proof of the second equality in Eq.~(\ref{eq:second}) follows from the
identities\footnote{The abelian version of Eq.~(\ref{eq:JSIDa}) was derived
  by Schwinger in Ref.~\cite{Schwinger:1951nm}. The nonabelian version was
  shown to KL long ago by Kim Milton.}
\bea\label{eq:JSIDa}
 e^{-i{\bs Q}\cdot {\bs \omega}}Q_a  e^{i{\bs Q}\cdot {\bs \omega}} &=&
\left( e^{i{\bs F}\cdot {\bs \omega}}\right)_{ab} Q_b \,;\\
 e^{-i{\bs Q}\cdot {\bs \omega}}\frac{\partial}{\partial \omega_a}
e^{i{\bs Q}\cdot {\bs \omega}} &=&
i\CG_{ab}({\bs \omega})Q_b \,.
\eea
These imply 
\be\label{eq:JSIDb}
\frac{\partial}{\partial \omega_a} e^{i{\bs Q}\cdot {\bs \omega}} =
i\CG_{ba}({\bs \omega}) Q_b  e^{i{\bs Q}\cdot {\bs \omega}}\,.
\ee
Hence,
\bea\label{eq:extrema}
&& \frac{\partial}{\partial \omega_a} \CH'({\bs \omega}) \equiv
\frac{\partial}{\partial \omega_a} \left(e^{i{\bs Q}\cdot {\bs \omega}}
\, \CH' \, e^{-i{\bs Q}\cdot {\bs \omega}}\right) = i\CG_{ba}({\bs \omega})
\bigl[Q_b, \CH'({\bs \omega})\bigr] \nn\\\\
\Lra &&\frac{\partial E(W)}{\partial \omega_a}\biggr|_{{\bs \omega} = {\bs
    \omega_0}} \equiv i \CG_{ba}({\bs \omega_0}) \lvac\bigl[Q_b, \CH'({\bs
    \omega_0})\bigr]\rvac = 0\,.\nn
\eea
Since $\CG({\bs {\omega}})$ is invertible, this implies $\lvac[Q_a, \CH'({\bs
    \omega_0})]\rvac = 0$. Differentiating again and using Eq.~(\ref{eq:Msq})
    gives the second half of Eq.~(\ref{eq:dtwoEz}):
\be\label{eq:ptwoE}
\frac{\partial^2 E(W_0)}{\partial \omega_{a,0} \partial \omega_{b,0}}
= \left(\CG(-{\bs \omega_0}) \,F_\pi^2 M^2\, \CG({\bs \omega_0})\right)_{ab} \,.
\ee

In deriving our formula, we ignored the singularities in
$d\omega_{a,0}/d\Lambda$ at phase transitions. This is not a problem at a
2-OPT where the zero in $M^2$ cancels the divergence in the derivatives. At a
1-OPT, $d\omega_{a,0}/d\Lambda$ and $d^2E(W_0)/d\Lambda^2$ are proportional
to $\delta$-functions, so the formula, while consistent, really has no
meaning there.

\section*{Appendix B: The CP Phase Transition in\\ Dashen's Three-Quark Model}

We illustrate Eq.~(\ref{eq:dtwoEz}) with the model Dashen discussed in
Ref.~\cite{Dashen:1971et}. Consider QCD with three massless quarks, $u,d,s$.
Their chiral flavor symmetry $G_f = SU(3)_L\otimes SU(3)_R$ is spontaneously
broken to $S_f = SU(3)_V$ in the vacuum $\rvac$ defined by
\be\label{eq:qcond}
\lvac\ol q_{Rj} q_{Li}\rvac = \lvac\ol q_{Li} q_{Rj}\rvac =
- \delta_{ij} \Delta_q \,,
\ee
where $\Delta_q \simeq 2\pi F_\pi^3$. The $G_f$-symmetry is also explicitly
broken by
\be\label{eq:CHprime}
\CH' = \ol q_R M_q q_L + \ol q_L M_q^\dag q_R \equiv \ol q M_q q
\ts,
\ee
where the (assumed) real quark mass matrix is
\be\label{eq:Mq}
 M_q = M_q^\dag \equiv
    \pm \left(\ba{ccc}
      m_u & 0 & 0\\
      0 & m_d & 0\\
      0 & 0 & m_s\\
      \ea\right)
= \pm m_s \left(\ba{ccc}
      \Lambda & 0 & 0\\
      0 & \Lambda & 0\\
      0 & 0 & 1\\
      \ea\right)
\ee
For simplicity, we assume isospin invariance, $m_u = m_d =m_s \Lambda$, with
$\Lambda \ge 0$.\footnote{The isospin-violating case was considered in
  Ref.~\cite{Creutz:2003xu}; also, K.~Lane and A.~O.~Martin, unpublished.
  While it has a considerably richer CP-phase diagram, it is easier to see
  the working of Eq.~(\ref{eq:dtwoEz}) in the isospin-conserving case.} This
Hamiltonian conserves CP.

The vacuum energy to be minimized is
\bea\label{eq:EW}
E(W=W_L W_R^\dag) &=&
\lvac\ol q_R (W_R^\dag M_q W_L) q_L
            + \ol q_L (W_L^\dag M_q W_R) q_R \rvac \nn\\
&=& -{\rm Tr}\left(M_q W + M_q^\dag W^\dag\right)\Delta_q \equiv
E(W^*) \,.
\eea
To minimize $E(W)$ with this mass matrix, we may restrict $W$ to the subspace
in which only $\omega_8$ is varied:
\be\label{eq:WformD}
W = e^{2it_8\omega_8} =  \left(\ba{ccc}
      e^{i\whomega} & 0 & 0\\
      0 & e^{i\whomega}  & 0\\
      0 & 0 & e^{-2i\whomega}\\
      \ea\right) \,,
\ee
where $\whomega =\omega_8/\sqrt{3}$. The vacuum energy is then
\be\label{eq:EWtwo}
E(W) = \mp 2\left[2\Lambda \cos\whomega +
\cos(2\whomega)\right] m_s\Delta_q \,.
\ee
The PGB masses are calculated from $F_\pi^2 M^2_{ab} = {\rm Tr}
[\{t_a,\{t_b, M_q(W+W^\dag)\}\}]\Delta_q$:
\bea\label{eq:PGBmass}
F_\pi^2 M_\pi^2 &\equiv& F_\pi^2 M_{33}^2 = \pm 4\Lambda
\cos\whomega \,m_s\Delta_q \,,\nn\\ 
F_\pi^2 M_K^2 &\equiv& F_\pi^2 M_{44}^2 = \pm 2\left[\Lambda
  \cos\whomega + \cos(2\whomega)\right] m_s \Delta_q \,,\nn\\
F_\pi^2 M_\eta^2 &\equiv&  F_\pi^2 M_{88}^2 = \pm
\textstyle{\frac{4}{3}}\left[\Lambda \cos\whomega + 
  2 \cos(2\whomega)\right] m_s\Delta_q \,.
\eea
For the plus sign in Eq.~(\ref{eq:Mq}), the strong interactions are in the
CPC phase with $\omega_{8,0} = 0$ and the minimizing matrix $W_0 = 1$; $E(W)
= -2(2\Lambda + 1) m_s\Delta_q$; and PGB masses $M_\pi^2 : M_K^2 : M_\eta^2 =
4\Lambda m_s: 2(\Lambda+ 1)m_s : 4/3(\Lambda + 2)m_s$.

The negative $M_q$ is more interesting. In this case $\ol \theta \equiv
\arg\det(M_q) = \pi$. When $\Lambda \ge 2$, the vacuum energy is minimized
for $\whomegaz = \pm\pi$; this is also a CPC phase.  When $0\le \Lambda < 2$,
the minimum occurs for $\cos\whomegaz = -\half \Lambda$, $\sin\whomegaz = \pm
\half \sqrt{4-\Lambda^2}$; this is a CPV phase. The phase $\whomegaz$ varies
from $\pm \pi/2$ to $\pm \pi$, with the two signs corresponding to the two
distinct CP-violating ground states. To summarize,
\be\label{eq:omeightD}
\whomegaz = 
\mp \tan^{-1}\left(\frac{\sqrt{4-\Lambda^2}}{\Lambda}\right) \theta(2-\Lambda) 
\pm \pi \, \theta(\Lambda-2) \,,
\ee
corresponding to
\bea\label{eq:WD}
 W_0 &&= \left(\ba{ccc}
       \half(-\Lambda \pm i\sqrt{4-\Lambda^2}) & 0 & 0\\
       0 &  \half(-\Lambda \pm i\sqrt{4-\Lambda^2}) & 0\\
       0 & 0 & -1 +\half \Lambda^2 \pm \frac{i}{2} \Lambda \sqrt{4-\Lambda^2}\\
       \ea\right) \,\theta(2-\Lambda)\nn \\\nn \\
     && + \left(\ba{rrr}
          -1 & 0 & 0\\
           0 & -1 & 0\\
           0 & 0 & 1\\
          \ea\right) \,\theta(\Lambda-2)\,.
\eea
Note the characteristic square-root singularity in the derivative of the
order parameter $\omega_{8,0}$. This is very similar to what we saw at the
2-OPTs in Figs.~\ref{fig:AGB_fig_8},\ref{fig:AGB_fig_9}. The vacuum energy is
\be\label{eq:EWD}
E(W_0) = -2\left[(1+\Lambda^2/2)\,\,\theta(2-\Lambda) +
  (2\Lambda-1)\,\,\theta(\Lambda-2) \right] ms\Delta_q\,,
\ee
and the PGB masses are
\bea\label{eq:PGBmassm}
F_\pi^2 M_\pi^2 &=&2\left[\Lambda^2 \,\theta(2-\Lambda) +
  2\Lambda\,\theta(\Lambda-2) \right] m_s\Delta_q \,,\nn\\ 
F_\pi^2 M_K^2 &=& 2\left[\,\theta(2-\Lambda) +
   (\Lambda-1)\,\theta(\Lambda-2)\right] m_s\Delta_q \,,\nn\\ 
F_\pi^2 M_\eta^2 &\equiv& F_\pi^2 M^2_{88} =
\textstyle{\frac{1}{3}}\left[2(4 - \Lambda^2)\,\theta(2-\Lambda) +
4(\Lambda-2)\,\theta(\Lambda-2)\right] m_s\Delta_q \,.
\eea
The $\Lambda = 2$ transition is second order, with $M^2_\eta \ra 0$
continuously there. The $\eta$ is this model's AGB. All the $F_\pi^2 M^2$ are
continuous at the transition, but their derivatives are not.

Finally, we demonstrate the equality in Eq.~(\ref{eq:dtwoEz}). It works
because $\CH'$ depends at most linearly on the parameter $\Lambda$. The
derivatives of the energy are
\bea\label{eq:LHS}
\frac{dE(W_0)}{d\Lambda} &=& -2\left[\Lambda\,\theta(2-\Lambda) +
  2\,\theta(\Lambda-2)\right] m_s\Delta_q \,,\\
\frac{d^2E(W_0)}{d\Lambda^2} &=& -2m_s\Delta_q\, \theta(2-\Lambda)\,.
\eea
The delta-function terms vanished. Note the discontinuity in the second
derivative at $\Lambda=2$. Since $\CG_{a8}(\whomegaz) = \delta_{a8}$, the
right-hand side of Eq.~(\ref{eq:dtwoEz}) is
\bea\label{eq:RHS}
&& -\left(\CG(-{\bs \omega_0}) \,
F_\pi^2 M^2\, \CG({\bs \omega_0})\right)_{ab} \frac{d\omega_{a,0}}{d\Lambda}
\frac{d\omega_{b,0}}{d\Lambda}
 =  -F_\pi^2 M^2_{88} \left(\frac{d\omega_{8,0}}{d\Lambda}\right)^2 \nn\\
&&\quad =  -\frac{m_s\Delta_q}{3}\left[2(4 - \Lambda^2)\,\theta(2-\Lambda) +
4(\Lambda-2)\,\theta(\Lambda-2)\right]\,
\left(\sqrt{\frac{3}{4-\Lambda^2}}\,\theta(2-\Lambda)\right)^2 \nn\\
&&\quad = -2 m_s\Delta_q \,\theta(2-\Lambda) =
\frac{d^2E(W_0)}{d\Lambda^2}\,.
\eea

\vfil\eject

\bibliography{AGB} \bibliographystyle{utcaps}
\end{document}